\documentclass[12pt, draftclsnofoot, onecolumn]{IEEEtran}
\usepackage{epsfig,graphicx,subfigure,psfrag,amsmath,cases}
\usepackage{latexsym,amssymb,amsmath,epsfig,subfigure,algorithm,mathtools}
\usepackage{algorithmic}
\usepackage{color}
\usepackage{url}
\usepackage{scrtime}
\author{\authorblockN{Yan Sun, Derrick Wing Kwan Ng, Zhiguo Ding, and Robert Schober\\
\thanks{Yan Sun and Robert Schober are  with the Institute for Digital Communications, Friedrich-Alexander-University Erlangen-N\"urnberg (FAU), Germany (email:\{yan.sun, robert.schober\}@fau.de). Derrick Wing Kwan Ng is with the School of Electrical Engineering and Telecommunications, the University of New South Wales, Australia (email: w.k.ng@unsw.edu.au). Zhiguo Ding is with the School of Computing and Communications, Lancaster University, United Kingdom (email: z.ding@lancaster.ac.uk). This paper has been accepted in part for presentation at IEEE Globecom 2016 \cite{sun2016optimal}.
}}
}

\title{Optimal Joint Power and Subcarrier Allocation for Full-Duplex Multicarrier Non-Orthogonal Multiple Access Systems}

\newtheorem{Thm}{Theorem}

\newtheorem{Def}{Definition}

\newtheorem{T-Prob}{Transformed Problem}

\DeclareMathOperator{\maxo}{maximize}
\DeclareMathOperator{\mino}{minimize}

 \newcommand{\qed}{\hfill \ensuremath{\blacksquare}}

\newcommand{\abs}[1]{\lvert#1\rvert}
\newcommand{\norm}[1]{\lVert#1\rVert}
\begin{document}
\maketitle \vspace*{-13mm}
\begin{abstract}
In this paper, we investigate resource allocation algorithm design for multicarrier non-orthogonal multiple access (MC-NOMA) systems employing a full-duplex (FD) base station (BS) for serving multiple half-duplex (HD) downlink (DL) and uplink (UL) users simultaneously. The proposed algorithm is obtained from the solution of a non-convex optimization problem for the maximization of the weighted sum system throughput. We apply monotonic optimization to develop an optimal joint power and subcarrier allocation policy. The optimal resource allocation policy serves as a system performance benchmark due to its high computational complexity.
Furthermore, a suboptimal iterative scheme based on successive convex approximation is proposed to strike a balance between computational complexity and optimality. Our simulation results reveal that the proposed suboptimal algorithm achieves a close-to-optimal performance.
Besides, FD MC-NOMA systems employing the proposed resource allocation algorithms provide a substantial system throughput improvement compared to conventional HD multicarrier orthogonal multiple access (MC-OMA) systems and other baseline schemes. Also, our results unveil that the proposed FD MC-NOMA systems achieve a fairer resource allocation compared to traditional HD MC-OMA systems.
\end{abstract}\vspace*{-0mm}
\vspace*{-2mm}
\section{Introduction}
Multicarrier multiple access techniques have been widely adopted in broadband wireless communication systems over the last decade, due to their flexibility in resource allocation and their ability to exploit multiuser diversity \cite{JR:Roger_OFDMA}\nocite{ng2012energy,Subpair}--\cite{Cui2009Distrib}. In conventional multicarrier systems, a given radio frequency band is divided into multiple orthogonal subcarriers and each subcarrier is allocated to at most one user to avoid multiuser interference (MUI). The spectral efficiency of such systems can be improved significantly by performing joint user scheduling and power allocation.
In \cite{ng2012energy}, the authors studied the resource allocation algorithm design for energy-efficient communication in multi-cell orthogonal frequency division multiple access (OFDM) systems.
In \cite{Subpair}, an asymptotically optimal power control and subcarrier allocation algorithm was proposed to maximize the transmission rate in OFDM systems with multiple relays.
The authors of \cite{Cui2009Distrib} proposed a distributed subcarrier, power, and rate allocation algorithm for the maximization of the weighted sum throughput of relay-assisted OFDM systems.
However, even with the schemes proposed in \cite{ng2012energy}\nocite{Subpair}--\cite{Cui2009Distrib}, the spectral resources are still underutilized as some subcarriers may be assigned exclusively to users with poor channel conditions to ensure fairness in resource allocation.

Non-orthogonal multiple access (NOMA) has recently received significant attention since it enables the multiplexing of multiple users simultaneously utilizing the same frequency resource, which improves system spectral efficiency \cite{ImpactPairNOMA}\nocite{ding2015general,minmaxNOMA,Zhang16Energy,saito2013non}--\cite{JointNOMA}. Since multiplexing multiple users on the same frequency channel leads to MUI, successive interference cancellation (SIC) is performed at the receivers to remove the undesired interference.
The authors of \cite{ImpactPairNOMA} investigated the impact of user pairing on the sum-rate of NOMA systems, and it was shown that the system throughput can be increased by pairing users enjoying good  channel conditions with users suffering from poor channel conditions.
In \cite{ding2015general}, a transmission framework based on signal alignment was proposed for multiple-input multiple-output (MIMO) NOMA systems.
A suboptimal joint power allocation and precoding design was presented in \cite{minmaxNOMA} for the maximization of the system throughput in multiuser MIMO-NOMA single-carrier systems.
In \cite{Zhang16Energy}, the optimal power allocation strategy for the maximization of the energy-efficiency of NOMA systems was investigated.
Yet, we note that \cite{ImpactPairNOMA}\nocite{ding2015general,minmaxNOMA}--\cite{Zhang16Energy}  focus only on the application of NOMA in single-carrier systems. In fact, spectral efficiency can be further improved by applying NOMA in multicarrier systems by exploiting the degrees of freedom offered by multiuser diversity and the power domain simultaneously.
In \cite{saito2013non}, the authors demonstrated that MC-NOMA systems employing a suboptimal power allocation scheme achieve system throughput gains over conventional multicarrier orthogonal multiple access (MC-OMA) systems.
The authors of \cite{JointNOMA} proposed a suboptimal joint power and subcarrier allocation algorithm for MC-NOMA systems.
Yet, the resource allocation schemes proposed in \cite{saito2013non,JointNOMA} are strictly suboptimal. Thus, the maximum achievable improvement in spectral efficiency of optimal MC-NOMA systems compared to MC-OMA systems is still unknown.
In our preliminary work \cite{sun2016optimal}, we studied the optimal joint power and subcarrier allocation algorithm design for maximization of the weighted sum throughput in MC-NOMA systems.
It was shown that MC-NOMA employing the proposed optimal resource allocation algorithm provided a substantial system throughput improvement compared to MC-OMA and the suboptimal scheme in \cite{JointNOMA}.
However, the radio spectral resources are not fully exploited in \cite{sun2016optimal}\nocite{JR:Roger_OFDMA,ng2012energy,Subpair,Cui2009Distrib,ImpactPairNOMA,ding2015general,minmaxNOMA,saito2013non}--\cite{JointNOMA},
since the base station (BS) operates in the half-duplex (HD) mode, where uplink (UL) and downlink (DL) transmissions employ orthogonal radio resources leading to spectrum underutilization.

Recently, full-duplex (FD) wireless communication has attracted significant research interest due to its potential to double the spectral efficiency by allowing simultaneous DL and UL transmission in the same frequency band \cite{FDRad}\nocite{FD_smlcll,ng2015power,Sun16FDSecurity,Li14Rate,Xiao16Achievable,Dynamic_FDrelay,Jiang15OFDMFD}--\cite{Nam15Joint}.
Therefore, it is expected that the spectral efficiency of traditional HD systems can be further improved by employing an FD BS.
In practice, the major challenges in FD communications are self-interference (SI) and the co-channel interference (CCI) between DL and UL users. In particular, the SI at the FD BS is caused by the signal leakage from the DL transmission to the UL signal reception, while the CCI is caused by the UL user signals interfering the DL users.
Several resource allocation designs for FD systems were proposed to overcome these challenges.
For example, in \cite{FD_smlcll}, a suboptimal DL beamformer was designed to improve the system throughput in FD MIMO systems.
In \cite{ng2015power}, the authors investigated simultaneous DL and UL transmission via an FD BS in distributed antenna systems and proposed an optimal joint power allocation and antenna selection algorithm minimizing the total network power consumption.
The tradeoff between the total DL and UL power consumption in FD systems was studied in \cite{Sun16FDSecurity}, where an optimal robust DL beamforming and UL power allocation algorithm was proposed to achieve power-efficient and secure communications.
In addition, the use of FD transceivers in multicarrier systems to improve spectral efficiency was also studied in \cite{Li14Rate}\nocite{Xiao16Achievable,Dynamic_FDrelay,Jiang15OFDMFD}--\cite{Nam15Joint}.
The rate region and the achievable sum rate of bidirectional communication links in a two-user FD OFDM system were studied in \cite{Li14Rate} and \cite{Xiao16Achievable}, respectively.
In \cite{Dynamic_FDrelay}, the authors proposed an optimal joint precoding and scheduling algorithm for the maximization of the weighted sum throughput in MIMO-OFDM-FD relaying systems.
In \cite{Jiang15OFDMFD}, a joint relay selection and subcarrier and power allocation algorithm for the maximization of the weighted sum throughput was proposed for multiuser FD-OFDM relaying systems.
The authors of \cite{Nam15Joint} studied a multiuser multicarrier network where an FD BS served multiple FD nodes and a joint subcarrier and power allocation algorithm for the maximization of the weighted sum throughput was proposed for the considered system.
However, only OMA schemes were considered for simultaneous DL and UL transmission in \cite{FDRad}\nocite{FD_smlcll,ng2015power,Sun16FDSecurity,Li14Rate,Xiao16Achievable,Dynamic_FDrelay,Jiang15OFDMFD}--\cite{Nam15Joint}, where orthogonality was achieved either in the spatial or in the frequency domain. However, the spectral efficiency can be further improved by incorporating NOMA into FD systems. In particular, for multicarrier systems, a new form of multiuser diversity can be exploited  by pairing multiple DL and UL users on each subcarrier.
However, a careful design of power allocation and user scheduling is vital for the performance of FD MC-NOMA systems due to the inherent interference.
To the best of our knowledge, FD MC-NOMA systems have not been investigated in the literature yet.
Thus, the achievable improvement in spectral efficiency of FD MC-NOMA systems compared to conventional HD-MC-OMA systems is unknown and the optimal resource allocation design for FD MC-NOMA systems has not been reported yet.

In this paper, we address the above issues. To this end,  we formulate the resource allocation algorithm design for the maximization of the weighted sum throughput of FD MC-NOMA systems as a non-convex optimization problem. The optimal power and subcarrier allocation policy can be obtained by solving the considered problem via monotonic optimization \cite{tuy2000monotonic}\nocite{zhang2013monotonic}--\cite{bjornson2013optimal}. Also, a low computational complexity suboptimal algorithm based on successive convex approximation is proposed  and shown to achieve a close-to-optimal performance. Our simulation results confirm the considerable improvement in spectral efficiency of the proposed FD MC-NOMA system compared to traditional HD MC-OMA systems. Besides, the results also indicate that a careful SI suppression is necessary to realize the potential performance gains introduced by FD MC-NOMA systems.

\vspace*{-5mm}
\section{System Model}
In this section, we present the adopted notation and the considered FD MC-NOMA system model.

\vspace*{-4mm}
\subsection{Notation}%
We use boldface lower case letters to denote vectors. $\mathbf{a}^T$ denotes the transpose of vector $\mathbf{a}$; $\mathbb{C}$ denotes the set of complex numbers; $\mathbb{R}$ denotes the set of non-negative real numbers; $\mathbb{R}^{N\times 1}$ denotes the set of all $N\times 1$ vectors with real entries and $\mathbb{R}^{N\times 1}_{\mathrm{+}}$ denotes the non-negative subset of $\mathbb{R}^{N\times 1}$; $\mathbb{Z}^{N\times 1}$ denotes the set of all $N\times 1$ vectors with integer entries; $\mathbf{a} \le \mathbf{b}$ indicates that $\mathbf{a}$ is component-wise smaller than $\mathbf{b}$; $\abs{\cdot}$ denotes the absolute value of a complex scalar; ${\cal E}\{\cdot\}$ denotes statistical expectation. The circularly symmetric complex Gaussian distribution with mean $w$ and variance $\sigma^2$ is denoted by ${\cal CN}(w,\sigma^2)$; and $\sim$ stands for ``distributed as". $\nabla_{\mathbf{x}} f(\mathbf{x})$ denotes the gradient vector of a function $f(\mathbf{x})$ whose components are the partial derivatives of $f(\mathbf{x})$.
\begin{figure}
\centering\vspace*{-3mm}
\includegraphics[width=5in]{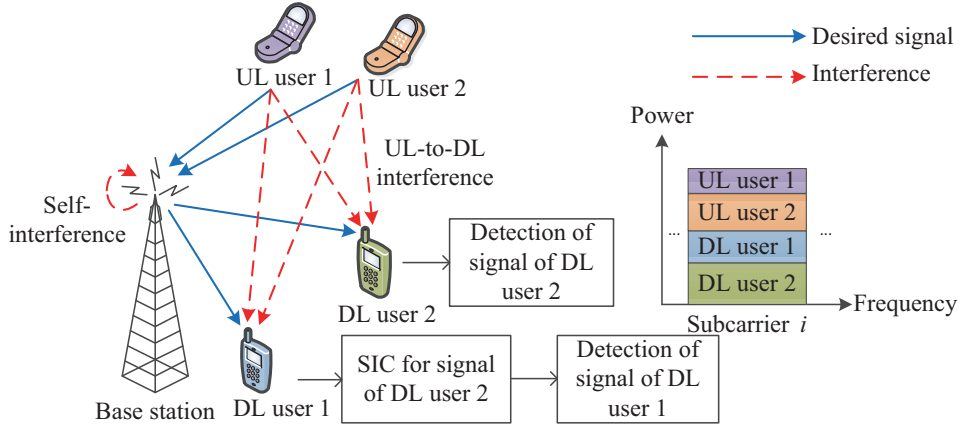}\vspace*{-3mm}
\caption{An FD MC-NOMA system where two DL users and two UL users are multiplexed on subcarrier $i$.
DL user $1$ decodes and removes DL user $2$'s signal before decoding its own desired signal. The FD BS first decodes UL user $1$'s signal and then decodes UL user $2$'s signal by removing UL user $1$'s interference signal.}
\label{fig:system_model}\vspace*{-5mm}
\end{figure}

\vspace*{-4mm}
\subsection{FD MC-NOMA System}

We consider an FD MC-NOMA system which comprises an FD BS, $K$ DL users, and $J$ UL users. All DL and UL users are equipped with a single antenna. Besides, the FD BS is also equipped with a single antenna for enabling simultaneous DL transmission and UL reception in the same frequency band\footnote{We note that FD radio prototypes equipped with a circulator can transmit and receive signals simultaneously with a single antenna \cite{FDRad} .}.
We assume that the BS and the DL users are equipped with successive interference cancellers, cf. Figure \ref{fig:system_model}. The entire frequency band of $W$ Hertz is partitioned into ${N_{\mathrm{F}}}$ orthogonal subcarriers.
In this paper, we assume that each subcarrier is allocated to at most two DL users and two UL users to limit the MUI and the UL-to-DL CCI on each subcarrier\footnote{The MUI and the UL-to-DL CCI per subcarrier increase as more DL and UL users are multiplexed on the same subcarrier which can degrade the performance of individual users.} and to ensure low hardware complexity and low processing delay\footnote{NOMA requires SIC at the receivers. In practice, a user performing SIC has to demodulate and decode the signals intended for other users in addition to its own signal. Thus, hardware complexity and processing delay increase with the number of users multiplexed on the same subcarrier \cite{saito2013non}.}.

Assuming DL user $m\in \{1,\ldots,K\}$, DL user $n\in \{1,\ldots,K\}$, UL user $r\in \{1,\ldots,J\}$, and UL user $t\in \{1,\ldots,J\}$  are selected and multiplexed on subcarrier $i \in \{1,\ldots,N_{\mathrm{F}}\}$, the received signals at DL user $m$, DL user $n$, and the BS are given by \begin{eqnarray}
\hspace*{-5mm} y_{\mathrm{DL}_m}^{i}\hspace*{-3mm}&=&\hspace*{-3mm}\sqrt{p_m^i \varpi_m}h_m^i x_{\mathrm{DL}_m}^{i} \hspace*{-0.5mm} + \hspace*{-0.5mm} \underbrace{\sqrt{p_n^i \varpi_m}h_m^i x_{\mathrm{DL}_n}^{i}}_{\mbox{MUI}} \hspace*{-0.5mm} +
\underbrace{\sqrt{q_r^i \vartheta_{r,m}} f_{r,m}^i x_{\mathrm{UL}_r}^{i} + \sqrt{q_t^i \vartheta_{t,m}}f_{t,m}^i x_{\mathrm{UL}_t}^{i}}_{\mbox{UL-to-DL CCI}} + z_{\mathrm{DL}_m}^{i},  \\
\hspace*{-5mm} y_{\mathrm{DL}_n}^{i}\hspace*{-3mm}&=&\hspace*{-3mm}\sqrt{p_n^i \varpi_n}h_n^i x_{\mathrm{DL}_n}^{i}  \hspace*{-0.5mm} + \hspace*{-0.5mm} \underbrace{\sqrt{p_m^i \varpi_n}h_n^i x_{\mathrm{DL}_m}^{i}}_{\mbox{MUI}} \hspace*{-0.5mm} +
\underbrace{\sqrt{q_r^i \vartheta_{r,n}} f_{r,n}^i x_{\mathrm{UL}_r}^{i} + \sqrt{q_t^i \vartheta_{t,n}}f_{t,n}^i x_{\mathrm{UL}_t}^{i}}_{\mbox{UL-to-DL CCI}} + z_{\mathrm{DL}_n}^{i},\,\, \text{and} \\
\hspace*{-5mm} y_{\mathrm{BS}}^{i}\hspace*{-3mm}&=&\hspace*{-3mm}\sqrt{q_r^i \varrho_r}g_r^i x_{\mathrm{UL}_r}^{i} + \sqrt{q_t^i \varrho_t}g_t^i x_{\mathrm{UL}_t}^{i} + \underbrace{l_{\mathrm{SI}}^i (\sqrt{p_m^i} x_{\mathrm{DL}_m}^{i} + \sqrt{p_m^i} x_{\mathrm{DL}_m}^{i})}_{\mbox{self-interference}} + z_{\mathrm{BS}}^{i},  \,\,\,\,
\end{eqnarray}
respectively. Variables $x_{\mathrm{DL}_m}^{i}\in\mathbb{C}$ and $x_{\mathrm{UL}_r}^{i}\in\mathbb{C}$ denote the symbols transmitted from the FD BS to DL user $m$ and from UL user $r$ to the FD BS on subcarrier $i$, respectively. Besides, without loss of generality, ${\cal E}\{\abs{x_{\mathrm{DL}_m}^{i}}^2\}={\cal E}\{\abs{x_{\mathrm{UL}_r}^{i}}^2\}=1, \forall m,r$ is assumed. $p_m^i$ is the transmit power of the signal intended for DL user $m$ at the FD BS and $q_r^i$ is the transmit power of the signal intended for the FD BS at UL user $r$ on subcarrier $i$. $h_m^i\in\mathbb{C}$, $g_r^i\in\mathbb{C}$, and $f_{r,m}^i\in\mathbb{C}$ denote the small scale fading coefficients for the link between the FD BS and DL user $m$, the link between UL user $r$ and the FD BS, and the link between UL user $r$ and DL user $m$ on subcarrier $i$, respectively.
$l_{\mathrm{SI}}^i\in\mathbb{C}$ denotes the SI channel at the FD BS.
Variables $\varpi_m\in\mathbb{R}$, $\varrho_r\in\mathbb{R}$, and $\vartheta_{r,m}\in\mathbb{R}$ represent the joint effect of path loss and shadowing between the FD BS and DL user $m$, between UL user $r$ and the FD BS, and between UL user $r$ and DL user $m$, respectively. $z_{\mathrm{DL}_m}^i\sim{\cal CN}(0,\sigma_{\mathrm{z}_{\mathrm{DL}_m}}^2)$ and $z_{\mathrm{BS}}^i\sim{\cal CN}(0,\sigma_{\mathrm{z}_{\mathrm{BS}}}^2)$ denote the complex additive white Gaussian noise (AWGN) on subcarrier $i$ at DL user $m$ and the FD BS, respectively. Besides, for the study of optimal resource allocation algorithm design, we assume that global channel state information (CSI) of all links in the network is available at the BS so as to unveil the performance upper bound of FD MC-NOMA systems.

\vspace*{-2mm}
\section{Resource Allocation Problem Formulation}
In this section, we first define the adopted performance measure for the considered FD MC-NOMA system. Then, we formulate the power and subcarrier allocation problem.

\vspace*{-2mm}
\subsection{Weighted System Throughput}
In the considered FD MC-NOMA system, a subcarrier can be allocated to at most two DL users and two UL users simultaneously.
In general, the power of the UL user signals is smaller than that of the signals emitted by the BS for DL users, which makes it difficult for the DL users to demodulate and remove the UL signal by performing SIC. Besides, different modulation and coding schemes may be utilized in the DL and the UL due to their different quality of service (QoS) requirements and different constraints on receiver hardware complexity \cite{holma2009lte}. Thus, in general, the DL users cannot demodulate and decode the UL signals. Therefore, in this paper, we assume that the DL users can only perform SIC to remove the signals of other DL users but treat all UL user signals as noise.
For illustration, we first assume a particular policy for subcarrier allocation and the SIC decoding order\footnote{The optimal  policy for subcarrier allocation and the SIC decoding order will be found by optimization in the next section.}. We assume that DL users $m, n$ and UL users $r,t$ are multiplexed on subcarrier $i$. Besides, DL user $n$ performs SIC to decode and remove DL user $m$'s signal. Also, the FD BS first decodes UL user $r$'s signal and then removes it by SIC before decoding UL user $t$'s signal. The weighted sum throughput on subcarrier $i$ under such a policy is given by
\begin{eqnarray} \label{throughput_i}
&&\hspace*{-5mm}U_{m,n,r,t}^i(\mathbf{p},\mathbf{q},\mathbf{s}) \notag\\[-1mm]
=&&\hspace*{-5mm}s_{m,n,r,t}^{i}\Bigg[
w_m  \log_2 \Big(1 +  \frac{H_{m}^i p_{m}^i } {H_{m}^i p_{n}^i  + F_{r,m}^i q_r^i + F_{t,m}^i q_t^i+ 1}\Big)
+\hspace*{-0mm}w_n\log_2 \Big(1 + \frac{H_{n}^i p_{n}^i}{F_{r,n}^i q_r^i + F_{t,n}^i q_t^i+ 1} \Big) \notag \\
+&&\hspace*{-5mm}\mu_r \log_2 \Big(1 +  \frac{G_{r}^i q_{r}^i } {G_{t}^i q_{t}^i + \rho L_{\mathrm{SI}}^i(p_m^i+p_n^i) + 1}\Big)
+\hspace*{-0mm}\mu_t \log_2 \Big(1 +  \frac{G_{t}^i q_{t}^i } {\rho L_{\mathrm{SI}}^i(p_m^i+p_n^i) + 1}\Big) \Bigg],
\end{eqnarray}
where $H_{m}^i=\frac{\varpi_m\abs{h_m^i}^2}{\sigma_{\mathrm{z}_{\mathrm{DL}_m}}^2}$, $G_{r}^i=\frac{\varrho_r\abs{g_r^i}^2}{\sigma_{\mathrm{z}_{\mathrm{BS}}}^2}$, $F_{r,m}^i=\frac{\vartheta_{r,m}\abs{f_{r,m}^i}^2}{\sigma_{\mathrm{z}_{\mathrm{DL}_m}}^2}$, and $L_{\mathrm{SI}}^i=\frac{\abs{l_{\mathrm{SI}}^i}^2}{\sigma_{\mathrm{z}_{\mathrm{BS}}}^2}$, respectively.
Variable $s_{m,n,r,t}^{i}\in\{0,1\}$ is the subcarrier allocation indicator. Specifically, $s_{m,n,r,t}^{i}=1$ if DL users $m$ and $n$ and UL users $r$ and $t$ are multiplexed on subcarrier $i$ where DL user $n$ performs SIC of DL user $m$'s signal and the FD BS first decodes UL user $r$'s signal and removes it before decoding UL user $t$'s signal. $s_{m,n,r,t}^{i}=0$ if another resource allocation policy is used.
The non-negative constants $0 \hspace*{-0mm} \le\hspace*{-0mm}  w_m \hspace*{-0mm}\le \hspace*{-0mm}1$ and $0  \le  \mu_r \le 1$ in \eqref{throughput_i} denote the priorities of DL user $m$ and UL user $r$ in resource allocation, respectively, which are specified in the media access control (MAC) layer to achieve a certain notion of fairness.
In practice, SI cannot be cancelled perfectly even if the SI channel is perfectly known at the FD BS due to the limited dynamic range of the receiver \cite{JR:FD_model}. Therefore, we model the residual SI after cancellation at the receive antenna as an independent zero-mean Gaussian distortion noise whose variance is proportional to the received power of the antenna \cite{JR:FD_model},
i.e., $\rho L_{\mathrm{SI}}^i(p_m^i+p_n^i)$ in \eqref{throughput_i}, where $0 <\rho \ll 1$ is a constant modelling the quality of the SI cancellation at the FD BS.

NOMA systems exploit the power domain for multiple access where different users are served at different power levels. In particular, for a given subcarrier, assume DL user $n$ desires to decode and remove the CCI caused by DL user $m$ via SIC. The interference cancellation is successful if user $n$'s received signal-to-interference-plus-noise ratio (SINR) for user $m$'s signal is larger than or equal to the received SINR of user $m$ for its own signal. In other words, DL user $n$ can only successfully decode and remove DL user $m$'s signal by SIC on subcarrier $i$ when the following inequality holds \cite{ImpactPairNOMA,tse2005fundamentals}:
\begin{eqnarray} \label{DL-inequ}
\log_2 \Big(1 + \frac{H_{n}^i p_{m}^i}{H_{n}^i p_{n}^i + F_{r,n}^i q_r^i + F_{t,n}^i q_t^i + 1} \Big) \ge
\log_2 \Big(1 + \frac{H_{m}^i p_{m}^i } {H_{m}^i p_{n}^i  + F_{r,m}^i q_r^i + F_{t,m}^i q_t^i+ 1}\Big).
\end{eqnarray}
The inequality in \eqref{DL-inequ} is equivalent to the following inequality:
\begin{eqnarray} \label{DL-inequ-constr}
Q_{m,n,r,t}^i(\mathbf{q}) \triangleq (H_n^i F_{r,m}^i - H_m^i F_{r,n}^i) q_r^i + (H_n^i F_{t,m}^i - H_m^i F_{t,n}^i) q_t^i + H_n^i -H_m^i \ge 0.
\end{eqnarray}
For facilitating the presentation, we denote $\mathbf{p}\in\mathbb{R}^{{N_{\mathrm{F}}}K \times 1}$, $\mathbf{q}\in\mathbb{R}^{{N_{\mathrm{F}}}J \times 1}$, and $\mathbf{s}\in\mathbb{Z}^{{N_{\mathrm{F}}}K^2 J^2 \times 1}$ as the collections of optimization variables $p_m^i$, $q_r^i$, and $s_{m,n,r,t}^i$, respectively.

We note that for the case of $m=n$ and $r=t$, the instantaneous weighted sum throughput on subcarrier $i$ in \eqref{throughput_i} becomes \vspace*{-2mm}
\begin{eqnarray} \label{OMA_throughput_i}
&& \hspace*{-5mm}U_{m,n,r,t}^i(\mathbf{p},\mathbf{q},\mathbf{s}) \notag \\
= && \hspace*{-5mm}s_{m,n,r,t}^{i}\Big[
w_m  \log_2 \Big(1 +  \frac{H_{m}^i (p_{m}^i + p_{n}^i)} {F_{r,m}^i q_r^i + F_{t,m}^i q_t^i+ 1}\Big)
+\mu_r \log_2 \Big(1 +  \frac{G_{r}^i (q_{r}^i + q_{t}^i)} {\rho L_{\mathrm{SI}}^i(p_m^i+p_n^i) + 1}\Big) \Big].
\end{eqnarray}
In fact, \eqref{OMA_throughput_i} is the instantaneous weighted throughput of subcarrier $i$ for FD MC-OMA, where $p_{m}^i +p_{n}^i, \forall m=n$ and $q_{r}^i + q_{t}^i, \forall r=t$ are the transmit powers allocated to DL user $m$ and UL user $r$ on subcarrier $i$, respectively. Therefore, \eqref{throughput_i}  generalizes the instantaneous weighted throughput of conventional FD MC-OMA systems to FD MC-NOMA systems.

\vspace*{-2mm}
\subsection{Optimization Problem Formulation}
The system objective is to maximize the weighted sum of the entire system throughput. The optimal joint power and subcarrier allocation policy is obtained by solving the following optimization problem\footnote{The proposed optimization framework can be extended to take into account a minimum required transmission rate for individual DL and UL users at the expense of a more involved notation.}:\vspace*{-0mm}
\begin{eqnarray} \label{pro}
\underset{\mathbf{p},\mathbf{q},\mathbf{s}}{\maxo} \hspace*{27mm}&&\hspace*{-30mm}\sum_{i=1}^{N_{\mathrm{F}}}\sum_{m=1}^{K} \sum_{n=1}^{K} \sum_{r=1}^{J} \sum_{t=1}^{J} U_{m,n,r,t}^i(\mathbf{p},\mathbf{q},\mathbf{s}) \notag  \\
\notag\mbox{s.t.}\,\,\,
\mbox{C1: }s_{m,n,r,t}^i Q_{m,n,r,t}^i(\mathbf{q}) \ge 0, \,\,\, \forall i,m,n,r,t, \quad
&&\hspace*{-6mm}\mbox{C2: }\overset{N_{\mathrm{F}}}{\underset{i=1}{\sum}} \overset{K}{\underset{m=1}{\sum}} \overset{K}{\underset{n=1}{\sum}} \overset{J}{\underset{r=1}{\sum}} \overset{J}{\underset{t=1}{\sum}} s_{m,n,r,t}^i (p_{m}^i \hspace*{-0.5mm}+\hspace*{-0.5mm} p_{n}^i) \hspace*{-0.5mm}\le\hspace*{-0.5mm} P_{\mathrm{max}}^{\mathrm{DL}}, \notag\\
\mbox{C3: }\overset{N_{\mathrm{F}}}{\underset{i=1}{\sum}} \overset{K}{\underset{m=1}{\sum}} \overset{K}{\underset{n=1}{\sum}} \overset{J}{\underset{t=1}{\sum}} s_{m,n,r,t}^i q_{r}^i \le P_{\mathrm{max}_r}^{\mathrm{UL}},\, \forall r, \quad \hspace*{0.3mm}
&&\hspace*{-6mm}\mbox{C4: } s_{m,n,r,t}^i \in \{0,1\},\,\,\ \forall i,m,n,r,t, \notag\\
\mbox{C5: } \overset{K}{\underset{m=1}{\sum}} \overset{K}{\underset{n=1}{\sum}} \overset{J}{\underset{r=1}{\sum}} \overset{J}{\underset{t=1}{\sum}} s_{m,n,r,t}^i \le 1,\,\, \forall i,\quad \hspace*{10.7mm}
&&\hspace*{-6mm}\mbox{C6: } p_m^i \ge 0, \,\, \forall i,m, \quad \hspace*{-0mm}\mbox{C7: } q_r^i \ge 0, \,\, \forall i,r.
\end{eqnarray}
Constraint C1 guarantees successful SIC at DL user $n$ if $s_{m,n,r,t}^{i}=1$.
We note that, for UL reception, since the FD BS is the receiver for all UL signals, it can perform SIC successfully in any desired order.
Constraint C2 is the power constraint for the BS with a maximum transmit power allowance of $P_{\mathrm{max}}^{\mathrm{DL}}$.
Constraint C3 limits the transmit power of UL user $r$ by $ P_{\mathrm{max}_r}^{\mathrm{UL}}$.
Constraints C4 and C5 are imposed to guarantee that each subcarrier is allocated to at most two DL users and two UL users.
Here, we note that DL user pairing, UL user pairing, and UL-to-DL user pairing are performed on each subcarrier.
Constraints C6 and C7 are the non-negative transmit power constraints for DL and UL users, respectively.

The problem in \eqref{pro} is a mixed combinatorial non-convex problem due to the binary constraint for subcarrier allocation in C4 and the non-convex objective function. In general, there is no systematic approach for solving mixed combinatorial non-convex problems. However, in the next section, we will exploit the monotonicity of the problem in \eqref{pro} to design an optimal resource allocation strategy for the considered system.

\vspace*{-2mm}
\section{Solution of the Optimization Problem}
In this section, we solve the problem in \eqref{pro} optimally by applying monotonic optimization. In addition, a suboptimal scheme is proposed which achieves a close-to-optimal performance at a low computational complexity.

\vspace*{-2mm}
\subsection{Monotonic Optimization}
First, we introduce some mathematical preliminaries of monotonic optimization \cite{tuy2000monotonic}\nocite{zhang2013monotonic}--\cite{bjornson2013optimal}.
\begin{Def}[Box]
Given any vector $\mathbf{z}\in\mathbb{R}^{N\times 1}_{\mathrm{+}}$, the hyper rectangle $[\mathbf{0},\mathbf{z}]=\{\mathbf{x}\mid \mathbf{0}\le\mathbf{x}\le\mathbf{z}\}$ is referred to as a box with vertex $\mathbf{z}$.
\end{Def}
\begin{Def}[Normal]
An infinite set $\mathcal{Z} \subset \mathbb{R}^{N\times 1}_{\mathrm{+}}$ is normal if given any element $\mathbf{z} \in \mathcal{Z}$, the box $[\mathbf{0},\mathbf{z}]\subset\mathcal{Z}$.
\end{Def}
\begin{Def}[Polyblock]
Given any finite set $\mathcal{V} \subset \mathbb{R}^{N\times 1}_{\mathrm{+}}$, the union of all boxes $[\mathbf{0},\mathbf{z}]$, $\mathbf{z}\in\mathcal{V}$, is a polyblock with vertex set $\mathcal{V}$.
\end{Def}
\begin{Def}[Projection]
Given any non-empty normal set $\mathcal{Z} \subset \mathbb{R}^{N\times 1}_{\mathrm{+}}$ and any vector $\mathbf{z}\in\mathbb{R}^{N\times 1}_{\mathrm{+}}$, $\Phi (\mathbf{z})$ is the projection of $\mathbf{z}$ onto the boundary of $\mathcal{Z}$, i.e.,  $\Phi\big(\mathbf{z}\big)=\lambda\mathbf{z}$, where $\lambda=\max\{\beta\mid \beta \mathbf{z} \in \mathcal{Z}\}$ and $\beta\in\mathbb{R}_{\mathrm{+}}$.
\end{Def}
\begin{Def} An optimization problem belongs to the class of monotonic optimization problems if it can be represented in the following form: \vspace*{-2mm}
    \begin{eqnarray} \label{MO}
    &&\hspace*{-10mm}\underset{\mathbf{z}}{\maxo} \,\,\,\Psi(\mathbf{z})\notag\\
    &&\hspace*{-10mm}\mbox{s.t.} \hspace*{7mm}\mathbf{z}\in\mathcal{Z},
    \end{eqnarray}
    where $\mathbf{z}$ is the vertex and set $\mathcal{Z}\subset \mathbb{R}^{N\times 1}_{\mathrm{+}}$ is a non-empty normal closed set and function $\Psi(\mathbf{z})$ is an increasing function on $\mathbb{R}^{N\times 1}_{\mathrm{+}}$.
\end{Def}

\vspace*{-2mm}
\subsection{Joint Power and Subcarrier Allocation Algorithm}
To facilitate the presentation of the optimal resource allocation algorithm,  we rewrite the weighted sum throughput of subcarrier $i$ in \eqref{throughput_i} in an equivalent form: \vspace*{-1mm}
\begin{eqnarray}
&&\hspace*{-6mm}U_{m,n,r,t}^i(\mathbf{p},\mathbf{q},\mathbf{s})\hspace*{-0.5mm} \\[-1mm]
= &&\hspace*{-6mm}\hspace*{-0.5mm}w_m \hspace*{-0.5mm} \log_2\hspace*{-0.5mm} \Big(1 \hspace*{-1mm}+ \hspace*{-1mm} \frac{s_{m,n,r,t}^{i} H_{m}^i p_{m}^i } {s_{m,n,r,t}^{i} (H_{m}^i p_{n}^i \hspace*{-0mm} + F_{r,m}^i q_r^i + F_{t,m}^i q_t^i) + 1}\Big) \hspace*{-1mm} +\hspace*{-0.5mm} w_n\log_2 \Big(1 \hspace*{-1mm}+ \hspace*{-1mm} \frac{s_{m,n,r,t}^{i} H_{n}^i p_{n}^i}{s_{m,n,r,t}^{i}(F_{r,n}^i q_r^i + F_{t,n}^i q_t^i) + 1} \Big) \notag\\
+ &&\hspace*{-6mm}\hspace*{-0.5mm}\mu_r \log_2 \Big(1 +  \frac{s_{m,n,r,t}^{i} G_{r}^i q_{r}^i } {s_{m,n,r,t}^{i} \big(G_{t}^i q_{t}^i + \rho L_{\mathrm{SI}}^i(p_m^i+p_n^i) \big)+ 1}\Big)
+\hspace*{-0mm}\mu_t \log_2 \Big(1 +  \frac{s_{m,n,r,t}^{i} G_{t}^i q_{t}^i } {s_{m,n,r,t}^{i} \rho L_{\mathrm{SI}}^i(p_m^i+p_n^i) + 1}\Big). \notag
\end{eqnarray}
Besides, we define $\tilde{p}_{m,n,r,t,m}^i=s_{m,n,r,t}^{i} p_{m}^i$, $\tilde{q}_{m,n,r,t,r}^i=s_{m,n,r,t}^{i} q_{r}^i$,
\begin{eqnarray}
u_{m,n,r,t}^i \hspace*{-2mm}&=&\hspace*{-2mm} 1 +  \frac{H_{m}^i \tilde{p}_{m,n,r,t,m}^i } {H_{m}^i \tilde{p}_{m,n,r,t,n}^i  + F_{r,m}^i \tilde{q}_{m,n,r,t,r}^i  +  F_{t,m}^i \tilde{q}_{m,n,r,t,t}^i  +  1},  \\
v_{m,n,r,t}^i \hspace*{-2mm}&=&\hspace*{-2mm} 1 +  \frac{ H_{n}^i \tilde{p}_{m,n,r,t,n}^i}{F_{r,n}^i \tilde{q}_{m,n,r,t,r}^i + F_{t,n}^i \tilde{q}_{m,n,r,t,t}^i + 1},\\
\zeta_{m,n,r,t}^i \hspace*{-2mm}&=&\hspace*{-2mm} 1 +  \frac{G_{r}^i \tilde{q}_{m,n,r,t,r}^i } {G_{t}^i \tilde{q}_{m,n,r,t,t}^i + \rho L_{\mathrm{SI}}^i(\tilde{p}_{m,n,r,t,m}^i+ \tilde{p}_{m,n,r,t,n}^i) + 1}, \,\,\,\, \text{and} \\
\xi_{m,n,r,t}^i \hspace*{-2mm}&=&\hspace*{-2mm} 1 +  \frac{ G_{t}^i \tilde{q}_{m,n,r,t,t}^i } { \rho L_{\mathrm{SI}}^i(\tilde{p}_{m,n,r,t,m}^i+\tilde{p}_{m,n,r,t,n}^i) + 1}.
\end{eqnarray}
Thus, the weighted sum throughput on subcarrier $i$ in \eqref{throughput_i} can be expressed as
\begin{eqnarray}
\hspace*{-3mm}U_{m,n,r,t}^i(\tilde{\mathbf{p}},\tilde{\mathbf{q}})\hspace*{1mm}
=\hspace*{1mm}\log_2(u_{m,n,r,t}^i)^{w_m}+ \log_2(v_{m,n,r,t}^i)^{w_n}+\log_2(\zeta_{m,n,r,t}^i)^{\mu_r}+\log_2(\xi_{m,n,r,t}^i)^{\mu_t},
\end{eqnarray}
where
$\tilde{\mathbf{p}}\in\mathbb{R}^{2N_{\mathrm{F}}K^2 J^2 \times1}$ is the collection of all $\tilde{p}_{m,n,r,t,m}^i$ and $\tilde{p}_{m,n,r,t,n}^i$, and $\tilde{\mathbf{q}}\in\mathbb{R}^{2N_{\mathrm{F}}K^2 J^2 \times1}$ is the collection of all $\tilde{q}_{m,n,r,t,r}^i$ and $\tilde{q}_{m,n,r,t,t}^i$.

Then, the original problem in \eqref{pro} can be rewritten as \vspace*{-2mm}
\begin{eqnarray} \label{eqv-pro}
\hspace*{-1mm}\underset{\tilde{\mathbf{p}},\tilde{\mathbf{q}},\mathbf{s}}{\maxo} \hspace*{70mm} && \hspace*{-77mm} \sum_{i=1}^{N_{\mathrm{F}}}\sum_{m=1}^{K} \sum_{n=1}^{K} \sum_{r=1}^{J} \sum_{t=1}^{J} \log_2(u_{m,n,r,t}^i)^{w_m} \hspace*{-1mm}+\hspace*{-1mm} \log_2(v_{m,n,r,t}^i)^{w_n} \hspace*{-1mm}+\hspace*{-1mm}  \log_2(\zeta_{m,n,r,t}^i)^{\mu_r} \hspace*{-1mm}+\hspace*{-1mm} \log_2(\xi_{m,n,r,t}^i)^{\mu_t} \notag \\
\mbox{s.t.} \,\,\,\, \mbox{C1: }Q_{m,n,r,t}^i(\tilde{\mathbf{q}}) \ge 0, \,\, \forall i,m,n,r,t, \quad \hspace*{18mm}
&&\hspace*{-17mm} \mbox{C2: }\overset{N_{\mathrm{F}}}{\underset{i=1}{\sum}} \overset{K}{\underset{m=1}{\sum}} \overset{K}{\underset{n=1}{\sum}} \overset{J}{\underset{r=1}{\sum}} \overset{J}{\underset{t=1}{\sum}} \, \tilde{p}_{m,n,r,t,m}^i\hspace*{-1mm}+\hspace*{-1mm}\tilde{p}_{m,n,r,t,n}^i \hspace*{-1mm}\le \hspace*{-1mm}P_{\mathrm{max}}^{\mathrm{DL}}, 
\notag \\[-1mm]
\mbox{C3: }\overset{N_{\mathrm{F}}}{\underset{i=1}{\sum}} \overset{K}{\underset{m=1}{\sum}} \overset{K}{\underset{n=1}{\sum}} \overset{J}{\underset{t=1}{\sum}} \tilde{q}_{m,n,r,t,r}^i \hspace*{-1mm}\le\hspace*{-1mm} P_{\mathrm{max}_r}^{\mathrm{UL}},\, \forall r,\quad \hspace*{9.8mm}
&&\hspace*{-17mm}  \mbox{C4, C5},  \notag\\
\mbox{C6: } \tilde{p}_{m,n,r,t,m}^i \ge 0, \,\, \forall i,m,n,r,t,
 \quad \hspace*{22mm}
&&\hspace*{-17mm}  \mbox{C7: } \tilde{q}_{m,n,r,t,r}^i \ge 0, \,\, \forall i,m,n,r,t,
\end{eqnarray}
where
\begin{eqnarray}
Q_{m,n,r,t}^i(\tilde{\mathbf{q}})
\hspace*{-2mm}&=&\hspace*{-2mm}(H_n^i F_{r,m}^i - H_m^i F_{r,n}^i) \tilde{q}_{m,n,r,t,r}^i + (H_n^i F_{t,m}^i - H_m^i F_{t,n}^i) \tilde{q}_{m,n,r,t,t}^i + H_n^i -H_m^i. \,\,
\end{eqnarray}

\begin{figure}[t]
 \centering\vspace*{-5mm}
\includegraphics[width=3in]{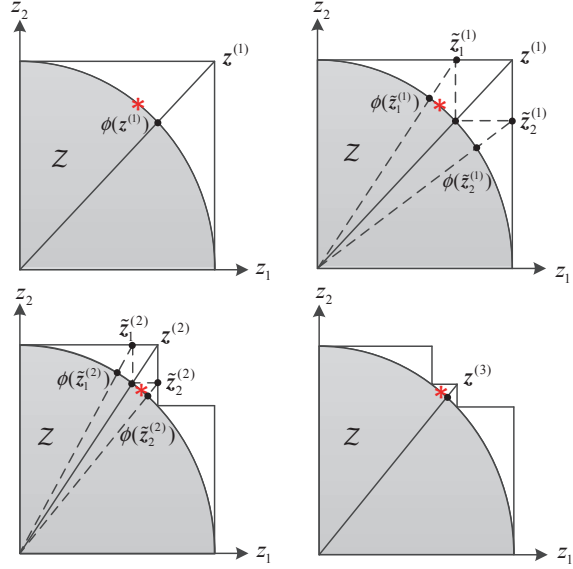} \vspace*{-2mm}
\caption{Illustration of the outer polyblock approximation algorithm for $D=2$. The red star is an optimal point on the boundary of the feasible set $\mathcal{Z}$.} \label{fig:polyblock}\vspace*{-5mm}
\end{figure}

Then, we define\vspace*{-2mm}
\begin{equation}
\hspace*{-6mm} f_{d}(\tilde{\mathbf{p}},\tilde{\mathbf{q}})\hspace*{-0mm}=
\hspace*{-0mm} \begin{cases}
\hspace*{-0mm} 1  +  H_{m}^i (\tilde{p}_{m,n,r,t,m}^i+\tilde{p}_{m,n,r,t,n}^i)  + F_{r,m}^i \tilde{q}_{m,n,r,t,r}^i  +  F_{t,m}^i \tilde{q}_{m,n,r,t,t}^i,
& d  = \Delta, \\
\hspace*{-0mm} 1+ H_{n}^i \tilde{p}_{m,n,r,t,n}^i + F_{r,n}^i \tilde{q}_{m,n,r,t,r}^i + F_{t,n}^i \tilde{q}_{m,n,r,t,t}^i ,
&  d= D/4 + \Delta, \\
\hspace*{-0mm} 1 + G_{r}^i (\tilde{q}_{m,n,r,t,r}^i + \tilde{q}_{m,n,r,t,t}^i) + \rho L_{\mathrm{SI}}^i(\tilde{p}_{m,n,r,t,m}^i+ \tilde{p}_{m,n,r,t,n}^i),
& d = D/2 + \Delta, \\
\hspace*{-0mm} 1+G_{t}^i \tilde{q}_{m,n,r,t,t}^i + \rho L_{\mathrm{SI}}^i(\tilde{p}_{m,n,r,t,m}^i+\tilde{p}_{m,n,r,t,n}^i), &  d = 3D/4 + \Delta,
\end{cases}
\end{equation}\vspace*{-1mm}
\begin{equation}
\hspace*{-22mm}g_{d}(\tilde{\mathbf{p}},\tilde{\mathbf{q}})\hspace*{-0mm}=
\hspace*{-0mm} \begin{cases}
\hspace*{-0mm} 1+H_{m}^i \tilde{p}_{m,n,r,t,n}^i  + F_{r,m}^i \tilde{q}_{m,n,r,t,r}^i  +  F_{t,m}^i \tilde{q}_{m,n,r,t,t}^i,  & d=\Delta, \\
\hspace*{-0mm} 1+F_{r,n}^i \tilde{q}_{m,n,r,t,r}^i + F_{t,n}^i \tilde{q}_{m,n,r,t,t}^i, & d=D/4+\Delta,\\
\hspace*{-0mm} 1+G_{t}^i \tilde{q}_{m,n,r,t,t}^i + \rho L_{\mathrm{SI}}^i(\tilde{p}_{m,n,r,t,m}^i+ \tilde{p}_{m,n,r,t,n}^i),  & d=D/2+\Delta, \\
\hspace*{-0mm} 1+\rho L_{\mathrm{SI}}^i(\tilde{p}_{m,n,r,t,m}^i+\tilde{p}_{m,n,r,t,n}^i), & d=3D/4+\Delta,
\end{cases}
\end{equation}
where $\Delta=(i-1)K^2 J^2+(m-1)K+(n-1)K+(r-1)J+t$ and $D=4N_{\mathrm{F}}K^2 J^2$.
In particular, functions $f_{d}(\tilde{\mathbf{p}},\tilde{\mathbf{q}})$ and $g_{d}(\tilde{\mathbf{p}},\tilde{\mathbf{q}})$ collect the numerator and denominator of variables $u_{m,n,r,t}^i$, $v_{m,n,r,t}^i$, $\zeta_{m,n,r,t}^i$, and $\xi_{m,n,r,t}^i$, respectively.
We further define $\mathbf{z}\hspace*{-1mm}=\hspace*{-1mm}[z_1\hspace*{-0.5mm},\hspace*{-0.5mm}\ldots\hspace*{-0.5mm},\hspace*{-0.5mm}z_D]^T\hspace*{-1.8mm}=\hspace*{-1mm}
[u_{1,1,1,1}^1\hspace*{-0.5mm}, \hspace*{-0.5mm}\ldots\hspace*{-0.5mm},\hspace*{-0.5mm}u_{K,K,J,J}^{N_{\mathrm{F}}}\hspace*{-0.3mm},
\hspace*{-0.3mm}v_{1,1,1,1}^1\hspace*{-0.5mm},\hspace*{-0.5mm}\ldots\hspace*{-0.5mm},\hspace*{-0.5mm}v_{K,K,J,J}^{N_{\mathrm{F}}},
\zeta_{1,1,1,1}^1\hspace*{-0.5mm}, \hspace*{-0.5mm}\ldots\hspace*{-0.5mm},\hspace*{-0.5mm}\zeta_{K,K,J,J}^{N_{\mathrm{F}}}\hspace*{-0.3mm},
\hspace*{-0.3mm}\xi_{1,1,1,1}^1\hspace*{-0.5mm},\hspace*{-0.5mm}\ldots\hspace*{-0.5mm},\hspace*{-0.5mm}\xi_{K,K,J,J}^{N_{\mathrm{F}}}]^T$.
Now, the original problem in $\eqref{pro}$ can be written as a standard monotonic optimization problem as: \vspace*{-3mm}
\begin{eqnarray}\label{MO-pro}
&&\hspace*{-15mm}\underset{\mathbf{z}}{\maxo}\,\, \,\, \notag \sum_{d=1}^{D} \log_2(z_d)^{\chi_d} \\
\mbox{s.t.}
&&\hspace*{-0mm} \mathbf{z}\in\mathcal{Z},
\end{eqnarray}
where $\hspace*{-0mm}\chi_d\hspace*{-0mm}$ is the equivalent user weight for $z_d$, i.e., $\chi_d\hspace*{-1mm}=\hspace*{-1mm}w_m$, $\forall d \in \{1,\ldots,D/4\}$, $\chi_d=w_n, \forall d \in \{D/4+1,\ldots,D/2\}$, $\chi_d=\mu_r, \forall d \in \{D/2+1,\ldots,3D/4\}$, and $\chi_d=\mu_t, \forall d \in \{3D/4+1,\ldots,D\}$. The feasible set $\mathcal{Z}$ is given by \vspace*{-3mm}
\begin{eqnarray}
\mathcal{Z}\hspace*{-1mm}=\Big\{ \mathbf{z} \mid 1\le z_d \le \frac{f_d(\tilde{\mathbf{p}},\tilde{\mathbf{q}})}{g_d(\tilde{\mathbf{p}},\tilde{\mathbf{q}})}, \,\,\, \tilde{\mathbf{p}},\tilde{\mathbf{q}}\in\mathcal{P}, \mathbf{s}\in\mathcal{S}, \forall d\Big\},
\end{eqnarray}
where $\mathcal{P}$ is the feasible set spanned by constraints $\mbox{C1--C3}$, $\mbox{C6}$, and $\mbox{C7}$, and $\mathcal{S}$ denotes the  feasible set spanned by constraints $\mbox{C4}$ and $\mbox{C5}$.
\begin{table}
\begin{algorithm} [H]                    
\caption{Outer Polyblock Approximation Algorithm}          
\label{alg1}                           
\begin{algorithmic} [1]
\small          
\STATE Initialize polyblock $\mathcal{B}^{{(1)}}$ with vertex set $\mathcal{V}^{{(1)}}=\{\mathbf{z}^{{(1)}}\}$ where the elements of $\mathbf{z}^{{(1)}}$ are set as \vspace*{-2mm}
    \begin{eqnarray}
    u_{m,n,r,t}^i=1+H_{m}^i P_{\mathrm{max}}^{\mathrm{DL}}, \,\,\,  v_{m,n,r,t}^i=1+H_{n}^i P_{\mathrm{max}}^{\mathrm{DL}},\,\,\, \zeta_{m,n,r,t}^i=1+G_{r}^i P_{\mathrm{max}_r}^{\mathrm{UL}}, \,\,\, \text{and} \,\,\, \xi_{m,n,r,t}^i=1+G_{t}^i P_{\mathrm{max}_t}^{\mathrm{UL}},\notag
    \end{eqnarray}
\STATE Set the error tolerance $\epsilon \ll 1$ and iteration index $k=1$

\REPEAT [Main Loop]
\STATE Construct a smaller polyblock $\mathcal{B}^{(k+1)}$ with vertex set $\mathcal{V}^{{(k+1)}}$ by replacing $\mathbf{z}^{{(k)}}$ with $D$ new vertices $\big\{\tilde{\mathbf{z}}^{{(k)}}_{1}, \ldots,\tilde{\mathbf{z}}^{{(k)}}_{D}\big\}$. The new vertex $\tilde{\mathbf{z}}^{{(k)}}_{d}$, $d\in\{1,\ldots,D\}$, is generated by \vspace*{-1mm}
        \begin{eqnarray}
        \tilde{\mathbf{z}}^{{(k)}}_{d}=\mathbf{z}^{{(k)}}-\Big(z^{{(k)}}_d-\phi_d\big(\mathbf{z}^{{(k)}}\big)\Big)\mathbf{e}_d, \notag
        \end{eqnarray}
        where $\phi_d\big(\mathbf{z}^{{(k)}}\big)$ is the $d$-th element of $\Phi\big(\mathbf{z}^{{(k)}}\big)$ which is obtained by $\textbf{Algorithm 2}$

\STATE  Find $\mathbf{z}^{{(k+1)}}$ as that vertex from $\mathcal{V}^{{(k+1)}}$ whose projection maximizes the objective function of the problem, i.e., \vspace*{-5mm}
        \begin{eqnarray}
        \mathbf{z}^{{(k+1)}}=\underset{\mathbf{z}\in\mathcal{V}^{{(k+1)}}}{\arg \max} \Big\{ \sum_{d=1}^{D} \log_2\big(\phi_d(\mathbf{z})\big)^{\chi_d}\Big\} \notag
        \end{eqnarray}

\STATE  $k=k+1$

\UNTIL $\frac{\norm{\mathbf{z}^{{(k)}}-\Phi(\mathbf{z}^{{(k)}})}}{\norm{\mathbf{z}^{{(k)}}}} \le \epsilon$
\STATE $\mathbf{z}^{*}=\Phi\big(\mathbf{z}^{{(k)}}\big)$ and $\{\tilde{\mathbf{p}}^*,\tilde{\mathbf{q}}^*\}$ is obtained when calculating $\Phi\big(\mathbf{z}^{{(k)}}\big)$
\end{algorithmic}
\end{algorithm}\vspace*{-14mm}
\end{table}
Now, we design a joint power and subcarrier allocation algorithm for solving the monotonic optimization problem in \eqref{MO-pro} based on the outer polyblock approximation approach \cite{tuy2000monotonic}\nocite{zhang2013monotonic}--\cite{bjornson2013optimal}. Since the objective function in \eqref{MO-pro} is a monotonic increasing function, the globally optimal solution is at the boundary of the feasible set $\mathcal{Z}$ \cite{tuy2000monotonic}\nocite{zhang2013monotonic}--\cite{bjornson2013optimal}. However, the boundary of $\mathcal{Z}$ is unknown. Therefore, we aim to approach the boundary by constructing a sequence of polyblocks. First, we construct a polyblock $\mathcal{B}^{{(1)}}$ that contains the feasible set $\mathcal{Z}$ with vertex set $\mathcal{V}^{{(1)}}$ which includes only one vertex $\mathbf{z}^{{(1)}}$. Then, we construct a smaller polyblock $\mathcal{B}^{{(2)}}$ based on $\mathcal{B}^{{(1)}}$ by replacing $\mathbf{z}^{{(1)}}$ with $D$ new vertices $\tilde{\mathcal{V}}^{{(1)}}=\big\{\tilde{\mathbf{z}}^{{(1)}}_{1}, \ldots,\tilde{\mathbf{z}}^{{(1)}}_{D}\big\}$. The feasible set $\mathcal{Z}$ is still contained in $\mathcal{B}^{{(2)}}$.
The new vertex $\tilde{\mathbf{z}}^{{(1)}}_{d}$ is generated as $\tilde{\mathbf{z}}^{{(1)}}_{d}=\mathbf{z}^{{(1)}}-\Big(z^{{(1)}}_d -\phi_d\big(\mathbf{z}^{{(1)}}\big)\Big)\mathbf{e}_d$, where  $\phi_d\big(\mathbf{z}^{{(1)}}\big)$ is the $d$-th element of $\Phi\big(\mathbf{z}^{{(1)}}\big)$, $\Phi\big(\mathbf{z}^{{(1)}}\big)\in \mathbb{C}^{D \times 1}$ is the projection of $\mathbf{z}^{{(1)}}$ on the feasible set $\mathcal{Z}$, and $\mathbf{e}_d$ is a unit vector that has a non-zero element only at index $d$.
Thus, the vertex set $\mathcal{V}^{{(2)}}$ of the newly generated polyblock $\mathcal{B}^{{(2)}}$ is $\mathcal{V}^{{(2)}}=(\mathcal{V}^{{(1)}}-\mathbf{z}^{{(1)}})\cup \tilde{\mathcal{V}}^{{(1)}}$. Then, we choose the optimal vertex from $\mathcal{V}^{{(2)}}$ whose projection maximizes the objective function of the problem in \eqref{MO-pro}, i.e., $\mathbf{z}^{{(2)}}=\underset{\mathbf{z}\in\mathcal{V}^{{(2)}}}{\arg \max} \Big\{ \sum_{d=1}^{D} \log_2\big(\phi_d(\mathbf{z})\big)^{\chi_d}\Big\}$. Similarly, we can repeat the above procedure to construct a smaller polyblock based on $\mathcal{B}^{{(2)}}$ and so on, i.e., $\mathcal{B}^{{(1)}} \supset \mathcal{B}^{{(2)}} \supset \dots \supset \mathcal{Z}$. The algorithm terminates if  $\frac{\norm{\mathbf{z}^{{(k)}}-\Phi(\mathbf{z}^{{(k)}})}}{\norm{\mathbf{z}^{{(k)}}}} \le \epsilon$, where $\epsilon > 0$ is the error tolerance which specifies the accuracy of the approximation. We illustrate the algorithm in Figure \ref{fig:polyblock} for $D=2$. The proposed outer polyblock approximation algorithm is summarized in \textbf{Algorithm 1}. In particular, the vertex $\mathbf{z}^{{(1)}}$ of the initial polyblock $\mathcal{B}^{{(1)}}$ is set by allocating on each subcarrier the maximum transmit powers $P_{\mathrm{max}}^{\mathrm{DL}}$ and $P_{\mathrm{max}_r}^{\mathrm{UL}}$ for all  DL  and all UL users, respectively, and omitting the MUI, the UL-to-DL CCI, and the SI. In fact, such intermediate resource allocation policy is infeasible in general. However, the corresponding polyblock contains the feasible set $\mathcal{Z}$ and the algorithm ultimately converges to one of the optimal points.

\textbf{Projection:} The projection of $\mathbf{z}^{{(k)}}$, i.e., $\Phi\big(\mathbf{z}^{{(k)}}\big)=\lambda\mathbf{z}^{{(k)}}$, in \textbf{Algorithm 1}, is obtained by solving \vspace*{-2mm}
\begin{eqnarray} \label{lambda}
\lambda\hspace*{-1mm}&=&\hspace*{-1mm}\max\{\beta\mid \beta \mathbf{z} \in \mathcal{Z}\} = \max\Big\{\beta\mid \beta \le \underset{1\le d \le D}{\min} \,\, \frac{f_d(\tilde{\mathbf{p}},\tilde{\mathbf{q}})}{z_d^{(k)} g_d(\tilde{\mathbf{p}},\tilde{\mathbf{q}})}, \,\,\, \tilde{\mathbf{p}},\tilde{\mathbf{q}}\in\mathcal{P}\Big\}\notag \\[-3mm]
\hspace*{-1mm}&=&\hspace*{-1mm} \underset{\tilde{\mathbf{p}},\tilde{\mathbf{q}}\in\mathcal{P}}{\max}\underset{1\le d \le D}{\min} \,\, \frac{f_d(\tilde{\mathbf{p}},\tilde{\mathbf{q}})}{z_d^{(k)} g_d(\tilde{\mathbf{p}},\tilde{\mathbf{q}})}.
\end{eqnarray}

\vspace*{-0mm}

\begin{table}[t]
\begin{algorithm} [H]                    
\caption{Projection Algorithm}          
\label{alg1}                           
\begin{algorithmic} [1]
\small          
\STATE Initialize $\lambda_1=0$

\STATE Set the error tolerance $\delta \ll 1$ and iteration index $n=1$

\REPEAT
\STATE $(\tilde{\mathbf{p}}_n^*,\tilde{\mathbf{q}}_n^*)=\underset{\tilde{\mathbf{p}},\tilde{\mathbf{q}}\in{\mathbf{\mathcal{P}}}}{\arg \max} \Big\{\underset{1\le d\le D}{\min}\big\{f_d(\tilde{\mathbf{p}},\tilde{\mathbf{q}})-\lambda_n z_d^{(k)}g_d (\tilde{\mathbf{p}},\tilde{\mathbf{q}})\big\}\Big\}$

\STATE $\lambda_{n+1}=\underset{1\le d \le D}{\min}\frac{f_d (\tilde{\mathbf{p}}_n^*,\tilde{\mathbf{q}}_n^*)}{z_d^{(k)}g_d (\tilde{\mathbf{p}}_n^*,\tilde{\mathbf{q}}_n^*)}$

\STATE $n=n+1$

\UNTIL $\underset{1\le d\le D}{\min}\big\{f_d(\tilde{\mathbf{p}}_{n-1}^*,\tilde{\mathbf{q}}_{n-1}^*)-\lambda_{n} z_d^{(k)}g_d (\tilde{\mathbf{p}}_{n-1}^*,\tilde{\mathbf{q}}_{n-1}^*)\big\} \le \delta$
\STATE The projection is $\Phi\big(\mathbf{z}^{{(k)}}\big)=\lambda_{n} \mathbf{z}^{{(k)}}$ and $\big\{\tilde{\mathbf{p}}_{n-1}^*, \tilde{\mathbf{q}}_{n-1}^*\big\}$ is the corresponding resource allocation policy.

\end{algorithmic}
\end{algorithm}\vspace*{-15mm}
\end{table}
The problem in \eqref{lambda} is a standard fractional programming problem which can be solved by the Dinkelbach algorithm \cite{JR:fractional} in polynomial time. The algorithm is summarized in \textbf{Algorithm 2}. Specifically, $\tilde{\mathbf{p}}_n^*$ and $\tilde{\mathbf{q}}_n^*$ in line 4 are obtained by solving the following convex optimization problem: \vspace*{-4mm}
\begin{eqnarray}\label{max-min-pro}
&&\hspace*{-5mm}(\tilde{\mathbf{p}}_n^*,\tilde{\mathbf{q}}_n^*)=
\underset{\tilde{\mathbf{p}},\tilde{\mathbf{q}}\in{\mathbf{\mathcal{P}}}}{\arg \max} \,\,\,\, \tau \notag \\[-2mm]
\mbox{s.t.}
&&\hspace*{-5mm} f_d(\tilde{\mathbf{p}},\tilde{\mathbf{q}})-\lambda_n z_d^{(k)}g_d (\tilde{\mathbf{p}},\tilde{\mathbf{q}}) \ge \tau, \,\, \forall d \in \{1,\ldots,D\},
\end{eqnarray}
where $\tau$ is an auxiliary variable. Hence, the power allocation policy is obtained when calculating the projection in \textbf{Algorithm 2}.
We note that the convex problem in \eqref{max-min-pro} can be solved by standard numerical solvers for solving convex programs such as CVX \cite{website:CVX}.

From the optimal vertex $\mathbf{z}^*$ obtained with \textbf{Algorithm 1}, we can obtain the optimal subcarrier allocation. In particular, we can restore the values of $u_{m,n,r,t}^i$, $v_{m,n,r,t}^i$, $\zeta_{m,n,r,t}^i$, and $\xi_{m,n,r,t}^i$ according to the mapping order of
\begin{eqnarray}
\hspace*{-10mm}\mathbf{z}\hspace*{-1mm}&=&\hspace*{-1mm}[z_1\hspace*{-0.5mm},\hspace*{-0.5mm}\ldots\hspace*{-0.5mm},\hspace*{-0.5mm}z_D]^T\hspace*{-1.8mm} \notag \\
\hspace*{-10mm}&=&\hspace*{-1mm}
[u_{1,1,1,1}^1\hspace*{-0.5mm}, \hspace*{-0.5mm}\ldots\hspace*{-0.5mm},\hspace*{-0.5mm}u_{K,K,J,J}^{N_{\mathrm{F}}}\hspace*{-0.3mm},
\hspace*{-0.3mm}v_{1,1,1,1}^1\hspace*{-0.5mm},\hspace*{-0.5mm}\ldots\hspace*{-0.5mm},\hspace*{-0.5mm}v_{K,K,J,J}^{N_{\mathrm{F}}},
\zeta_{1,1,1,1}^1\hspace*{-0.5mm}, \hspace*{-0.5mm}\ldots\hspace*{-0.5mm},\hspace*{-0.5mm}\zeta_{K,K,J,J}^{N_{\mathrm{F}}}\hspace*{-0.3mm},
\hspace*{-0.3mm}\xi_{1,1,1,1}^1\hspace*{-0.5mm},\hspace*{-0.5mm}\ldots\hspace*{-0.5mm},\hspace*{-0.5mm}\xi_{K,K,J,J}^{N_{\mathrm{F}}}]^T.
\end{eqnarray}
Besides, since $u_{m,n,r,t}^i$, $v_{m,n,r,t}^i$, $\zeta_{m,n,r,t}^i$, and $\xi_{m,n,r,t}^i$ are larger than one if DL users $m$ and $n$ and UL user $r$ and $t$ are scheduled on subcarrier $i$, we can obtain the optimal subcarrier allocation policy $\mathbf{s}^*$ as \vspace*{-4mm}
\begin{equation}
s_{m,n,r,t}^{i}\hspace*{-0mm}=
\hspace*{-0mm} \begin{cases}
\hspace*{-0mm} 1 & \hspace*{-0mm} u_{m,n,r,t}^i >1, v_{m,n,r,t}^i >1, \zeta_{m,n,r,t}^i >1, \,\,\text{and} \,\, \xi_{m,n,r,t}^i >1,\\[-2mm]
\hspace*{-0mm} 0 & \hspace*{-0mm} \text{otherwise}.
\end{cases}
\end{equation}

The proposed monotonic optimization based resource allocation algorithm provides a systematic procedure to achieve one of the globally optimal solutions in a finite number of  iterations. However, its computational complexity grows exponentially with the number of vertices, $D$, adopted in each iteration. Yet, the performance achieved by the optimal algorithm can serve as a performance upper bound for suboptimal algorithms. In the following, we propose a suboptimal resource allocation algorithm which has a polynomial time computational complexity to strike a balance between complexity and system performance.

\vspace*{-3mm}
\subsection{Suboptimal Solution}
In this section, we propose a suboptimal scheme with low computational complexity,  which obtains a locally optimal solution for the optimization problem in \eqref{pro}. Since \eqref{eqv-pro} is equivalent to \eqref{pro}, we focus on the solution of the problem in \eqref{eqv-pro}. We note that the product terms $\tilde{p}_{m,n,r,t,m}^i=s_{m,n,r,t}^i p_{m}^i$ and $\tilde{q}_{m,n,r,t,r}^i=s_{m,n,r,t}^i q_{r}^i$ in \eqref{eqv-pro} are the obstacles for the design of a computationally efficient resource allocation algorithm. In order to circumvent this difficulty, we adopt the big-M formulation to decompose the product terms \cite{lee2011mixed}. In particular, we impose the following additional constraints:\vspace*{-2mm}
\begin{eqnarray}
\hspace*{-9mm}\mbox{C8: } \tilde{p}_{m,n,r,t,m}^i \le P_{\mathrm{max}}^{\mathrm{DL}} s_{m,n,r,t}^i, \,\, \forall i,m,n,r,t, \hspace*{19.5mm} \quad &&\hspace*{-6mm}\mbox{C9: } \tilde{p}_{m,n,r,t,m}^i \le p_m^i, \,\, \forall i,m,n,r,t,   \\[-0.5mm]
\hspace*{-7mm}\hspace*{-4mm}\mbox{C10: } \tilde{p}_{m,n,r,t,m}^i \ge p_m^i \hspace*{-1mm} - \hspace*{-1mm}(1 \hspace*{-1mm} - \hspace*{-0mm}s_{m,n,r,t}^i)P_{\mathrm{max}}^{\mathrm{DL}} ,\,\, \forall i,m,n,r,t,
\quad &&\hspace*{-6mm} \mbox{C11: }  \tilde{p}_{m,n,r,t,m}^i \ge 0, \,\, \forall i,m,n,r,t, \\[-0.5mm]
\hspace*{-7mm}\mbox{C12: } \tilde{q}_{m,n,r,t,r}^i \le P_{\mathrm{max}_r}^{\mathrm{UL}} s_{m,n,r,t}^i, \,\, \forall i,m,n,r,t, \hspace*{17.5mm} \quad &&\hspace*{-6mm}\mbox{C13: } \tilde{q}_{m,n,r,t,r}^i \le q_r^i, \,\, \forall i,m,n,r,t,   \\[-0.5mm]
\hspace*{-7mm}\hspace*{-4mm}\mbox{C14: } \tilde{q}_{m,n,r,t,r}^i \ge q_r^i \hspace*{-1mm} - \hspace*{-1mm}(1 \hspace*{-1mm} - \hspace*{-0mm}s_{m,n,r,t}^i)P_{\mathrm{max}_r}^{\mathrm{UL}} ,\,\, \forall i,m,n,r,t, \hspace*{1.5mm}
\quad &&\hspace*{-6mm} \mbox{C15: }  \tilde{q}_{m,n,r,t,r}^i \ge 0, \,\, \forall i,m,n,r,t.
\end{eqnarray}
Besides, the integer constraint C4 in optimization problem \eqref{eqv-pro} is a non-convex constraint. Thus, we rewrite constraint C4 in the equivalent form: \vspace*{-1mm}
\begin{eqnarray}
&&\hspace*{-10mm}\text{C4}\mbox{a: } \overset{{N_{\mathrm{F}}}}{\underset{i=1}{\sum}} \overset{K}{\underset{m=1}{\sum}} \overset{K}{\underset{n=1}{\sum}} \overset{J}{\underset{r=1}{\sum}} \overset{J}{\underset{t=1}{\sum}} s_{m,n,r,t}^i - (s_{m,n,r,t}^i)^2 \le 0 \quad \text{and} \\
&&\hspace*{-10mm}\text{C4}\mbox{b: } 0 \le s_{m,n,r,t}^i \le 1,\,\, \forall i,m,n,r,t.
\end{eqnarray}
Now, optimization variables $s_{m,n,r,t}^i$ are continuous values between zero and one. Thus, we can reformulate the optimization problem in \eqref{eqv-pro} in the following equivalent form: \vspace*{-0mm}
\begin{eqnarray} \label{subopt-pro}
\hspace*{-5mm}&&\hspace*{-5mm}\underset{\tilde{\mathbf{p}},\tilde{\mathbf{q}},\mathbf{p},\mathbf{q},\mathbf{s}}{\mino}\,\, \,\, \notag \sum_{i=1}^{{N_{\mathrm{F}}}}\sum_{m=1}^{K} \sum_{n=1}^{K} \overset{J}{\underset{r=1}{\sum}} \overset{J}{\underset{t=1}{\sum}} -U_{m,n,r,t}^i(\tilde{\mathbf{p}},\tilde{\mathbf{q}})\notag \\
\hspace*{-1mm}&&\hspace*{4mm}\mbox{s.t.}  \hspace*{7mm}\,\,\mbox{C1--C3}, \text{C4}\mbox{a},\text{C4}\mbox{b}, \mbox{C5--C15}.
\end{eqnarray}

The non-convexity in \eqref{subopt-pro}  is caused by both the objective function and constraint $\mbox{C4a}$. In fact, constraint $\mbox{C4a}$ is the difference of two convex functions which is known as a reverse convex function \cite{ng2015power,che2014Joint,dinh2010local}. Hence, we introduce the following theorem for handling constraint $\mbox{C4a}$.
\begin{Thm}\label{thm-penalty}
For a sufficiently large constant value $\eta \gg 1$, the optimization problem in \eqref{subopt-pro} is equivalent to the following problem: \vspace*{-0mm}
    \begin{eqnarray} \label{penalty-pro}
    \hspace*{-5mm}&&\hspace*{-5mm}\underset{\tilde{\mathbf{p}},\tilde{\mathbf{q}},\mathbf{p},\mathbf{q},\mathbf{s}}{\mino}\,\, \,\, \notag \sum_{i=1}^{{N_{\mathrm{F}}}}\sum_{m=1}^{K} \sum_{n=1}^{K} \overset{J}{\underset{r=1}{\sum}} \overset{J}{\underset{t=1}{\sum}} -U_{m,n,r,t}^i(\tilde{\mathbf{p}},\tilde{\mathbf{q}}) + \eta\Big(\overset{{N_{\mathrm{F}}}}{\underset{i=1}{\sum}} \overset{K}{\underset{m=1}{\sum}} \overset{K}{\underset{n=1}{\sum}} \overset{J}{\underset{r=1}{\sum}} \overset{J}{\underset{t=1}{\sum}} s_{m,n,r,t}^i - (s_{m,n,r,t}^i)^2\Big) \notag \\
    \hspace*{-1mm}&&\hspace*{4mm}\mbox{s.t.}  \hspace*{7mm}\,\,\mbox{C1--C3}, \text{C4}\mbox{b}, \mbox{C5--C15},
    \end{eqnarray}
where $\eta$ acts as a penalty factor to penalize the objective function for any $s_{m,n,r,t}^i$ that is not equal to $0$ or $1$.
\end{Thm}

\emph{\quad Proof: } Please refer to the appendix. \hfill\qed

The resulting optimization problem in \eqref{penalty-pro} is still non-convex because of the objective function. To facilitate the presentation, we rewrite the problem as \vspace*{-0mm}
\begin{eqnarray}\label{dc-penalty-pro}
\hspace*{-1mm}&&\hspace*{-0mm}\underset{\tilde{\mathbf{p}},\tilde{\mathbf{q}},\mathbf{p},\mathbf{q},\mathbf{s}}{\mino}\,\, \,\, F(\tilde{\mathbf{p}},\tilde{\mathbf{q}})-G(\tilde{\mathbf{p}},\tilde{\mathbf{q}})+\eta(H(\mathbf{s})-M(\mathbf{s})) \notag \\
\hspace*{-1mm}&&\hspace*{4mm}\mbox{s.t.}  \hspace*{7mm}\,\,\mbox{C1--C3}, \text{C4}\mbox{b}, \mbox{C5--C15},
\end{eqnarray}
where \vspace*{-2mm}
\begin{eqnarray}
\hspace*{-0mm}F(\tilde{\mathbf{p}},\tilde{\mathbf{q}})
\hspace*{-3mm}&=&\hspace*{-3mm}\sum_{i=1}^{{N_{\mathrm{F}}}}\sum_{m=1}^{K} \sum_{n=1}^{K} \overset{J}{\underset{r=1}{\sum}} \overset{J}{\underset{t=1}{\sum}} \hspace*{-1mm} - \hspace*{-1mm} w_m \hspace*{-0.5mm} \log_2\hspace*{-1mm} \big(1 \hspace*{-1mm}+ \hspace*{-1mm} H_{m}^i (\tilde{p}_{m,n,r,t,m}^i \hspace*{-1mm} + \hspace*{-1mm} \tilde{p}_{m,n,r,t,n}^i) \hspace*{-1mm} + \hspace*{-1mm} F_{r,m}^i \tilde{q}_{m,n,r,t,r}^i  \hspace*{-1mm} + \hspace*{-1mm}  F_{t,m}^i \tilde{q}_{m,n,r,t,t}^i  \big)  \hspace*{-1mm} \notag \\
\hspace*{-5mm}&-&\hspace*{-3mm} w_n\log_2(1 \hspace*{-0mm}+ \hspace*{-0mm}H_{n}^i \tilde{p}_{m,n,r,t,n}^i + F_{r,n}^i \tilde{q}_{m,n,r,t,r}^i + F_{t,n}^i \tilde{q}_{m,n,r,t,t}^i) \notag \\
\hspace*{-5mm}&-&\hspace*{-3mm} \mu_r\log_2\big(1 \hspace*{-0mm}+ \hspace*{-0mm} G_{r}^i \tilde{q}_{m,n,r,t,r}^i + G_{t}^i \tilde{q}_{m,n,r,t,t}^i + \rho L_{\mathrm{SI}}^i(\tilde{p}_{m,n,r,t,m}^i+ \tilde{p}_{m,n,r,t,n}^i) \big) \notag \\
\hspace*{-5mm}&-&\hspace*{-3mm} \mu_t\log_2 \big(1 \hspace*{-0mm}+ \hspace*{0mm} G_{t}^i \tilde{q}_{m,n,r,t,t}^i + \rho L_{\mathrm{SI}}^i(\tilde{p}_{m,n,r,t,m}^i+\tilde{p}_{m,n,r,t,n}^i) \big) ,
\end{eqnarray}\vspace*{-6mm}
\begin{eqnarray}
\hspace*{-15mm} H(\mathbf{s}) \hspace*{-0mm}=\hspace*{-0mm} \overset{{N_{\mathrm{F}}}}{\underset{i=1}{\sum}} \overset{K}{\underset{m=1}{\sum}} \overset{K}{\underset{n=1}{\sum}} \overset{J}{\underset{r=1}{\sum}} \overset{J}{\underset{t=1}{\sum}} s_{m,n,r,t}^i, \,\,\,\, \,\,\,\,
M(\mathbf{s}) \hspace*{-0mm}=\hspace*{-0mm}\overset{{N_{\mathrm{F}}}}{\underset{i=1}{\sum}} \overset{K}{\underset{m=1}{\sum}} \overset{K}{\underset{n=1}{\sum}} \overset{J}{\underset{r=1}{\sum}} \overset{J}{\underset{t=1}{\sum}} (s_{m,n,r,t}^i)^2, \,  \text{and}
\end{eqnarray}
\begin{eqnarray}
\hspace*{-12mm}G(\tilde{\mathbf{p}},\tilde{\mathbf{q}})\hspace*{-3mm}
&=&\hspace*{-3mm}\sum_{i=1}^{{N_{\mathrm{F}}}}\sum_{m=1}^{K} \sum_{n=1}^{K} \overset{J}{\underset{r=1}{\sum}} \overset{J}{\underset{t=1}{\sum}} -w_m \hspace*{-0.5mm} \log_2\hspace*{0mm} (1+ H_{m}^i \tilde{p}_{m,n,r,t,n}^i  + F_{r,m}^i \tilde{q}_{m,n,r,t,r}^i  +  F_{t,m}^i \tilde{q}_{m,n,r,t,t}^i )\notag \\
\hspace*{-5mm}&-&\hspace*{-3mm} w_n \hspace*{-0.5mm} \log_2\hspace*{0mm} (1+F_{r,n}^i \tilde{q}_{m,n,r,t,r}^i + F_{t,n}^i \tilde{q}_{m,n,r,t,t}^i) \notag \\
\hspace*{-5mm}&-&\hspace*{-3mm} \mu_r \hspace*{-0.5mm} \log_2\hspace*{0mm} \big(1+ G_{t}^i \tilde{q}_{m,n,r,t,t}^i + \rho L_{\mathrm{SI}}^i(\tilde{p}_{m,n,r,t,m}^i+ \tilde{p}_{m,n,r,t,n}^i) \big)\notag \\
\hspace*{-5mm}&-&\hspace*{-3mm}\mu_t \hspace*{-0.5mm} \log_2\hspace*{0mm} \big(1+ \rho L_{\mathrm{SI}}^i(\tilde{p}_{m,n,r,t,m}^i+\tilde{p}_{m,n,r,t,n}^i) \big).
\end{eqnarray}\vspace*{-4mm}

We note that $F(\tilde{\mathbf{p}},\tilde{\mathbf{q}})$, $G(\tilde{\mathbf{p}},\tilde{\mathbf{q}})$, $H(\mathbf{s})$, and $M(\mathbf{s})$ are convex functions and the problem in \eqref{dc-penalty-pro} belongs to the class of difference of convex (d.c.) function programming. As a result, we can apply successive convex approximation \cite{dinh2010local} to obtain a locally optimal solution of \eqref{dc-penalty-pro}.
Since $G(\tilde{\mathbf{p}},\tilde{\mathbf{q}})$ and $M(\mathbf{s})$ are differentiable convex functions, for any feasible point $\tilde{\mathbf{p}}^{(k)}$, $\tilde{\mathbf{q}}^{(k)}$, and $\mathbf{s}^{(k)}$, we have the following inequalities:
\begin{eqnarray}\label{ineq1}
G(\tilde{\mathbf{p}},\tilde{\mathbf{q}}) \hspace*{-3mm}&\ge&\hspace*{-3mm} G(\tilde{\mathbf{p}}^{(k)},\tilde{\mathbf{q}}^{(k)}) \hspace*{-0.7mm}+\hspace*{-0.7mm}\nabla_{\tilde{\mathbf{p}}} G(\tilde{\mathbf{p}}^{(k)},\tilde{\mathbf{q}}^{(k)})^T(\tilde{\mathbf{p}}-\tilde{\mathbf{p}}^{(k)})
\hspace*{-0.7mm}+\hspace*{-0.7mm} \nabla_{\tilde{\mathbf{q}}} G(\tilde{\mathbf{p}}^{(k)},\tilde{\mathbf{q}}^{(k)})^T(\tilde{\mathbf{q}}-\tilde{\mathbf{q}}^{(k)})\,\,\, \text{and}\\
\label{ineq2}M(\mathbf{s}) \hspace*{-3mm}&\ge&\hspace*{-3mm} M(\mathbf{s}^{(k)}) +\nabla_{\mathbf{s}} M(\mathbf{s}^{(k)})^T(\mathbf{s}-\mathbf{s}^{(k)}),
\end{eqnarray}
where the right hand sides of \eqref{ineq1} and \eqref{ineq2} are affine functions representing the global underestimation of $G(\tilde{\mathbf{p}},\tilde{\mathbf{q}})$ and $M(\mathbf{s})$, respectively.
\begin{table}
 \vspace*{-5mm}
\begin{algorithm} [H]                  
\caption{Successive Convex Approximation}          
\label{alg1}                           
\begin{algorithmic} [1]
\small          
\STATE Initialize the maximum number of iterations $I_{\mathrm{max}}$, penalty factor $\eta \gg 1$, iteration index $k=1$, and initial point $\tilde{\mathbf{p}}^{(1)}$, $\tilde{\mathbf{q}}^{(1)}$, and $\mathbf{s}^{(1)}$

\REPEAT
\STATE Solve \eqref{dc} for a given $\tilde{\mathbf{p}}^{(k)}$, $\tilde{\mathbf{q}}^{(k)}$, and $\mathbf{s}^{(k)}$ and store the intermediate resource allocation policy $\{\tilde{\mathbf{p}}, \tilde{\mathbf{q}}, \mathbf{s}\}$

\STATE Set $k=k+1$ and $\tilde{\mathbf{p}}^{(k)}=\tilde{\mathbf{p}}$, $\tilde{\mathbf{q}}^{(k)}=\tilde{\mathbf{q}}$, and $\mathbf{s}^{(k)}=\mathbf{s}$

\UNTIL convergence or $k=I_{\mathrm{max}}$

\STATE $\tilde{\mathbf{p}}^{*}=\tilde{\mathbf{p}}^{(k)}$, $\tilde{\mathbf{q}}^{*}=\tilde{\mathbf{q}}^{(k)}$, and $\mathbf{s}^{*}=\mathbf{s}^{(k)}$

\end{algorithmic}
\end{algorithm}\vspace*{-15mm}
\end{table}
Therefore, for any given $\tilde{\mathbf{p}}^{(k)}$, $\tilde{\mathbf{q}}^{(k)}$, and $\mathbf{s}^{(k)}$, we can obtain an upper bound for \eqref{dc-penalty-pro} by solving the following convex optimization problem:
\begin{eqnarray}\label{dc}
\hspace*{-1mm}&&\hspace*{-0mm}\underset{\tilde{\mathbf{p}},\tilde{\mathbf{q}},\mathbf{p},\mathbf{q},\mathbf{s}}{\mino}\,\, \,\, F(\tilde{\mathbf{p}},\tilde{\mathbf{q}})-G(\tilde{\mathbf{p}}^{(k)},\tilde{\mathbf{q}}^{(k)}) -\nabla_{\tilde{\mathbf{p}}} G(\tilde{\mathbf{p}}^{(k)},\tilde{\mathbf{q}}^{(k)})^T(\tilde{\mathbf{p}}-\tilde{\mathbf{p}}^{(k)})  \notag \\
\hspace*{-1mm}&&\hspace*{14mm}-\hspace*{-0.7mm} \nabla_{\tilde{\mathbf{q}}} G(\tilde{\mathbf{p}}^{(k)},\tilde{\mathbf{q}}^{(k)})^T(\tilde{\mathbf{q}}-\tilde{\mathbf{q}}^{(k)})
\hspace*{-0.7mm}+\hspace*{-0.7mm}\eta\big(H(\mathbf{s})-M(\mathbf{s}^{(k)}) \hspace*{-0.7mm}-\hspace*{-0.7mm} \nabla_{\mathbf{s}} M(\mathbf{s}^{(k)})^T(\mathbf{s}-\mathbf{s}^{(k)})\big) \notag \\
\hspace*{-1mm}&&\hspace*{4mm}\mbox{s.t.}  \hspace*{7mm}\,\,\mbox{C1--C3}, \text{C4}\mbox{b}, \mbox{C5--C15},
\end{eqnarray}
where \vspace*{-3mm}
\begin{eqnarray}
\hspace*{-13mm}&&\hspace*{-3mm} \nabla_{\tilde{\mathbf{p}}}  G(\tilde{\mathbf{p}}^{(k)},\tilde{\mathbf{q}}^{(k)} )^T (\tilde{\mathbf{p}} - \tilde{\mathbf{p}}^{(k)} ) \notag \\
\hspace*{-13mm}&=&\hspace*{-3mm}\overset{{N_{\mathrm{F}}}}{\underset{i=1}{\sum}} \overset{K}{\underset{m=1}{\sum}}\overset{K}{\underset{n=1}{\sum}} \overset{J}{\underset{r=1}{\sum}}\overset{J}{\underset{t=1}{\sum}}  - \frac{w_m  H_m^i  (\tilde{p}_{m,n,r,t,n}^{i}  -  \tilde{p}_{m,n,r,t,n}^{i(k)})} {(1+ H_{m}^i \tilde{p}_{m,n,r,t,n}^{i(k)} + F_{r,m}^i \tilde{q}_{m,n,r,t,r}^{i(k)}  +  F_{t,m}^i \tilde{q}_{m,n,r,t,t}^{i(k)})\ln2} \notag \\
\hspace*{-13mm}&-&\hspace*{-3mm} \frac{\mu_r \rho L_{\mathrm{SI}}^i(\tilde{p}_{m,n,r,t,m}^i-\tilde{p}_{m,n,r,t,m}^{i(k)})+\mu_r \rho L_{\mathrm{SI}}^i (\tilde{p}_{m,n,r,t,n}^i-\tilde{p}_{m,n,r,t,n}^{i(k)})}{\big(1+ G_{t}^i \tilde{q}_{m,n,r,t,t}^{i(k)} + \rho L_{\mathrm{SI}}^i(\tilde{p}_{m,n,r,t,m}^{i(k)}+ \tilde{p}_{m,n,r,t,n}^{i(k)})\big)\ln2} \notag\\
\hspace*{-13mm}&-&\hspace*{-3mm} \frac{\mu_t \rho L_{\mathrm{SI}}^i (\tilde{p}_{m,n,r,t,m}^i-\tilde{p}_{m,n,r,t,m}^{i(k)}) + \mu_t \rho L_{\mathrm{SI}}^i (\tilde{p}_{m,n,r,t,n}^i-\tilde{p}_{m,n,r,t,n}^{i(k)}) }{\big( 1+ \rho L_{\mathrm{SI}}^i(\tilde{p}_{m,n,r,t,m}^{i(k)}+\tilde{p}_{m,n,r,t,n}^{i(k)}) \big)\ln2},
\end{eqnarray}
\begin{eqnarray}
\hspace*{-3mm}&&\hspace*{-3mm} \nabla_{\tilde{\mathbf{q}}}  G( \tilde{\mathbf{p}}^{(k)},\tilde{\mathbf{q}}^{(k)})^T  (\tilde{\mathbf{q}} - \tilde{\mathbf{q}}^{(k)} ) \notag \\
&=&\hspace*{-3mm}\overset{{N_{\mathrm{F}}}}{\underset{i=1}{\sum}} \overset{K}{\underset{m=1}{\sum}}\overset{K}{\underset{n=1}{\sum}} \overset{J}{\underset{r=1}{\sum}}\overset{J}{\underset{t=1}{\sum}}
- \frac{w_m  F_{r,m}^i (\tilde{q}_{m,n,r,t,r}^i - \tilde{q}_{m,n,r,t,r}^{i(k)})+  w_m F_{t,m}^i (\tilde{q}_{m,n,r,t,t}^i - \tilde{q}_{m,n,r,t,t}^{i(k)})}{(1+ H_{m}^i \tilde{p}_{m,n,r,t,n}^{i(k)}  + F_{r,m}^i \tilde{q}_{m,n,r,t,r}^{i(k)}  +  F_{t,m}^i \tilde{q}_{m,n,r,t,t}^{i(k)} )\ln2} \notag\\
\hspace*{-3mm}&-&\hspace*{-3mm} \frac{w_n F_{r,n}^i (\tilde{q}_{m,n,r,t,r}^i - \tilde{q}_{m,n,r,t,r}^{i(k)}) + w_n F_{t,n}^i (\tilde{q}_{m,n,r,t,t}^i - \tilde{q}_{m,n,r,t,t}^{i(k)})}{(1+F_{r,n}^i \tilde{q}_{m,n,r,t,r}^{i(k)} + F_{t,n}^i \tilde{q}_{m,n,r,t,t}^{i(k)}) \ln2} \notag \\
\hspace*{-3mm}&-&\hspace*{-3mm} \frac{\mu_r G_{t}^i (\tilde{q}_{m,n,r,t,t}^i-\tilde{q}_{m,n,r,t,t}^{i(k)})}{\big(1+ G_{t}^i \tilde{q}_{m,n,r,t,t}^{i(k)} + \rho L_{\mathrm{SI}}^i(\tilde{p}_{m,n,r,t,m}^{i(k)}+ \tilde{p}_{m,n,r,t,n}^{i(k)}) \big) \ln2}, \,\,\, \text{and}
\end{eqnarray}
\begin{eqnarray}
\hspace*{-12.5mm} \nabla_{{\mathbf{s}}} M( \mathbf{s}^{(k)} \hspace*{-0.5mm} )^T  (\mathbf{s} - \mathbf{s}^{(k)}\hspace*{-0.5mm} )
=\overset{{N_{\mathrm{F}}}}{\underset{i=1}{\sum}} \overset{K}{\underset{m=1}{\sum}} \overset{K}{\underset{n=1}{\sum}} \overset{J}{\underset{r=1}{\sum}}\overset{J}{\underset{t=1}{\sum}} 2s_{m,n,r,t}^{i (k)}(s_{m,n,r,t}^{i}-s_{m,n,r,t}^{i (k)}).
\end{eqnarray}
Then, we employ an iterative algorithm to tighten the obtained upper bound as summarized in \textbf{Algorithm 3}.
In each iteration, the convex problem in \eqref{dc} can be solved efficiently by standard convex program solvers such as CVX \cite{website:CVX}.
By solving the convex upper bound problem in \eqref{dc}, the proposed iterative scheme generates a sequence of feasible solutions $\tilde{\mathbf{p}}^{(k+1)}$, $\tilde{\mathbf{q}}^{(k+1)}$, and $\mathbf{s}^{(k+1)}$ successively. The proposed suboptimal iterative algorithm converges to a locally optimal solution of \eqref{dc} with a polynomial time computational complexity \cite{dinh2010local}.

\vspace*{-3mm}
\section{Simulation Results}

\begin{table}[t]\vspace*{-2mm}\caption{System parameters used in simulations.}\vspace*{-2mm}\label{tab:parameters} 
\newcommand{\tabincell}[2]{\begin{tabular}{@{}#1@{}}#2\end{tabular}}
\centering
\begin{tabular}{|l|l|}\hline
\hspace*{-1mm}Carrier center frequency and system bandwidth & $2.5$ GHz and  $5$ MHz \\
\hline
\hspace*{-1mm}The number of subcarriers, ${N_{\mathrm{F}}}$, and the bandwidth of each subcarrier & $64$ and  $78$ kHz \\
\hline
\hspace*{-1mm}Path loss exponent and SI cancellation constant, $\rho$ &  \mbox{$3.6$} and  \mbox{$-90$ dB}   \\
\hline
\hspace*{-1mm}DL user noise power and UL BS noise power, $\sigma_{\mathrm{z}_{{\mathrm{DL}_m}}}^2$ and $\sigma_{\mathrm{z}_{{\mathrm{BS}}}}^2$ &  \mbox{$-125$ dBm} and \mbox{$-125$ dBm}  \\
\hline
\hspace*{-1mm}Maximum transmit power for UL users, $P_{\mathrm{max}_r}^{\mathrm{UL}}$, and BS antenna gain &  \mbox{$18$ dBm} and \mbox{$10$ dBi}   \\
\hline
\hspace*{-1mm}The error tolerance $\delta$ for {\bf{Algorithm 1}} &  $0.01$  \\
\hline
\end{tabular}
\vspace*{-7mm}
\end{table}
\vspace*{-1mm}

In this section, we investigate the performance of the proposed resource allocation scheme through simulations.
We adopt the simulation parameters given in Table \ref{tab:parameters}, unless  specified otherwise.
A single cell with two ring-shaped boundary regions is considered. The outer boundary and the inner boundary have radii of $30$ meters and $600$ meters, respectively. The $K$ DL and $J$ UL users are randomly and uniformly distributed between the inner and the outer boundary.
The BS is located at the center of the cell.
The maximum transmit power of the FD BS is $P_{\mathrm{max}}^{\mathrm{DL}}$.
For the weight of the users, we choose the normalized distance between the users and the BS, i.e., $w_m=\frac{a_m}{\underset{i\in\{1,\ldots,K\}}{\max}\{a_i\}}$ and $\mu_r=\frac{b_r}{\underset{i\in\{1,\ldots,J\}}{\max}\{b_i\}}$, where $a_m$ and $b_r$ are the distances from DL user $m$ and UL user $r$ to the FD BS, respectively\footnote{The weights are chosen to provide resource allocation fairness, especially for the cell edge users which suffer from poor channel conditions. }.
The penalty term $\eta$ for the proposed suboptimal algorithm is set to $10 \log_2(1+\frac{P_{\mathrm{max}}^{\mathrm{DL}}}{\sigma_{\mathrm{z}_{{\mathrm{DL}}_m}}^2})$.
The small-scale fading of the DL channels, the UL channels, and the channel between the DL and UL users are modeled as independent and identically distributed Rayleigh fading. The fading coefficients of the SI channel on each subcarrier are generated as independent and identically distributed Rician random variables with Rician factor $5$ dB. The results shown in this section were averaged over different realizations of both path loss and multipath fading.

For comparison, we also consider the performance of three baseline schemes. For baseline scheme $1$, we consider an FD MC-OMA system where an FD BS communicates with at most one DL user and one UL user simultaneously on each subcarrier. In this case, the MUI between the DL users and the MUI between the UL users is avoided at the DL users and the FD BS, respectively. In particular, we set $m=n$ and $r=t$ and the utility function in \eqref{throughput_i} becomes \eqref{OMA_throughput_i}. Then, we can jointly optimize $\mathbf{s}$, $\mathbf{p}$, and $\mathbf{q}$ under the proposed optimization framework subject to the same set of constraints as in \eqref{pro}. We note that baseline scheme $1$ is actually a special case of the proposed optimal resource allocation scheme.
For baseline scheme $2$, we consider a HD MC-NOMA system where a HD BS performs DL transmission and UL reception in two orthogonal time intervals having equal durations. As a result, both the SI and the CCI between the DL and UL users are avoided and the HD BS can communicate with at most two DL users or two UL users on each subcarrier in either one of the time intervals.
In particular, for the DL transmission in baseline scheme $2$, we adopt the optimal resource allocation algorithm in \cite{sun2016optimal} to obtain the optimal power and subcarrier allocation policy. For the UL transmission, by following \cite{sun2016optimal}, we can formulate a joint UL power and subcarrier allocation problem  which is given by \vspace*{-2mm}
\begin{eqnarray} \label{pro-UL}
&&\hspace*{-1mm}\underset{q_{r}^i\ge 0, s_{r,t}^i}{\maxo}\,\, \,\, \sum_{i=1}^{N_{\mathrm{F}}}\sum_{r=1}^{J} \sum_{t=1}^{J} s_{r,t}^{i}\Big[\mu_r \log_2 \Big(1 +  \frac{G_{r}^i q_{r}^i } {G_{t}^i q_{t}^i + 1}\Big)
+\hspace*{-0mm}\mu_t \log_2 \Big(1 +  G_{t}^i q_{t}^i \Big)\Big]\\[-1mm]
\notag\mbox{s.t.}
&&\hspace*{-0mm}\mbox{C1: }\overset{N_{\mathrm{F}}}{\underset{i=1}{\sum}} \overset{J}{\underset{t=1}{\sum}} s_{r,t}^i q_{r}^i \le P_{\mathrm{max}_r}^{\mathrm{UL}},\, \forall r, \notag \quad \mbox{C2: } s_{r,t}^i \in \{0,1\},\,\,\ \forall i,r,t,  \quad
\mbox{C3: } \overset{J}{\underset{r=1}{\sum}} \overset{J}{\underset{t=1}{\sum}} s_{r,t}^i \le 1,\,\, \forall i, \quad
\end{eqnarray}
where $s_{r,t}^i$ is the subcarrier allocation indicator. The optimal joint power and subcarrier allocation policy for the UL transmission in baseline scheme $2$ is obtained by solving the problem in \eqref{pro-UL} following a similar approach as \cite{sun2016optimal}.
For baseline scheme $3$, we consider a traditional HD MC-OMA system. The joint power and subcarrier allocation for the DL and the UL transmissions are obtained by utilizing the algorithms in \cite{JR:Time_sharing_wei_yu} and \cite{kim2005joint}, respectively.
In order to have a fair comparison, the resulting throughputs for baseline schemes $2$ and $3$ are divided by two since either UL or DL transmission is performed at a given time.
The setting of the proposed schemes and the baseline schemes are summarized in Table \ref{tab:schemes}.

\vspace*{-2mm}
\subsection{Convergence of Proposed Algorithms}

\begin{figure}[t]
\makeatletter\def\@captype{table}\makeatother
\begin{minipage}[b]{0.4\linewidth}\caption{Setting of different schemes.}
\centering \label{tab:schemes} \small
\begin{tabular}{|l|l|l|}\hline
\hspace*{-1mm}Scheme & FD & NOMA \\
\hline
\hspace*{-1mm}Proposed schemes & Yes & Yes \\
\hline
\hspace*{-1mm}Baseline scheme 1& Yes & No \\
\hline
\hspace*{-1mm}Baseline scheme 2 & No & Yes \\
\hline
\hspace*{-1mm}Baseline scheme 3 & No & No \\
\hline
\end{tabular}
\end{minipage}
\makeatletter\def\@captype{figure}\makeatother
\begin{minipage}[c]{0.55\linewidth}
\centering\vspace*{-5mm}
\includegraphics[width=4in]{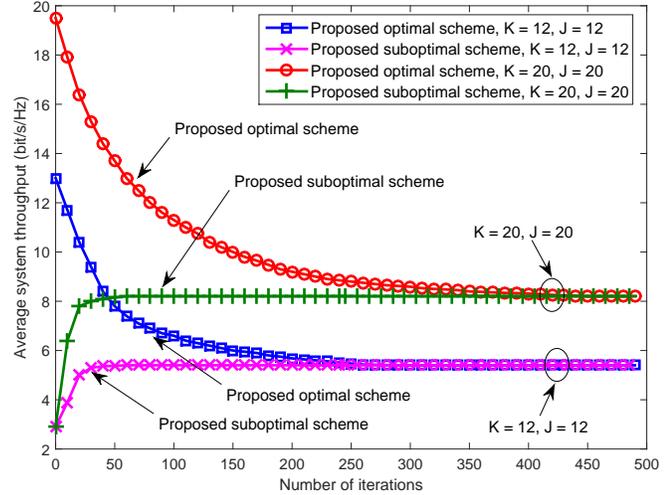}\vspace*{-5mm}
\caption{Convergence of the proposed optimal and suboptimal algorithms for different numbers of DL and UL users.}
\label{fig:throughput_vs_iterations}\vspace*{-8mm}
\end{minipage}
\end{figure}

Figure \ref{fig:throughput_vs_iterations} illustrates the convergence of the proposed optimal and suboptimal algorithms for different numbers of DL users, $K$, and UL users, $J$. The maximum transmit power for the FD BS is set to $P_{\mathrm{max}}^{\mathrm{DL}}=32 \text{ dBm}$. As can be seen from Figure \ref{fig:throughput_vs_iterations}, the proposed optimal and suboptimal algorithms both converge to the optimal solution for  different values of $K$ and $J$. Besides, we notice that the rate of  convergence of the proposed suboptimal algorithm is significantly faster than that of the proposed optimal algorithm. In particular, for $K=12$ and $J=12$, the proposed optimal algorithm converges to the optimal solution in less than $300$  iterations on average and the proposed suboptimal algorithm converges to a stationary point after $30$ iterations on average. For $K=20$ and $J=20$, the proposed optimal algorithm needs considerably more iterations to converge since additional users lead to additional search dimensions in the feasible solution set.
On the other hand, for the proposed suboptimal algorithm, the increase in the number of iterations is very small.

\vspace*{-4mm}
\subsection{Average System Throughput versus Maximum Transmit Power}
\begin{figure}
\centering\vspace*{-5mm}
\includegraphics[width=4.3in]{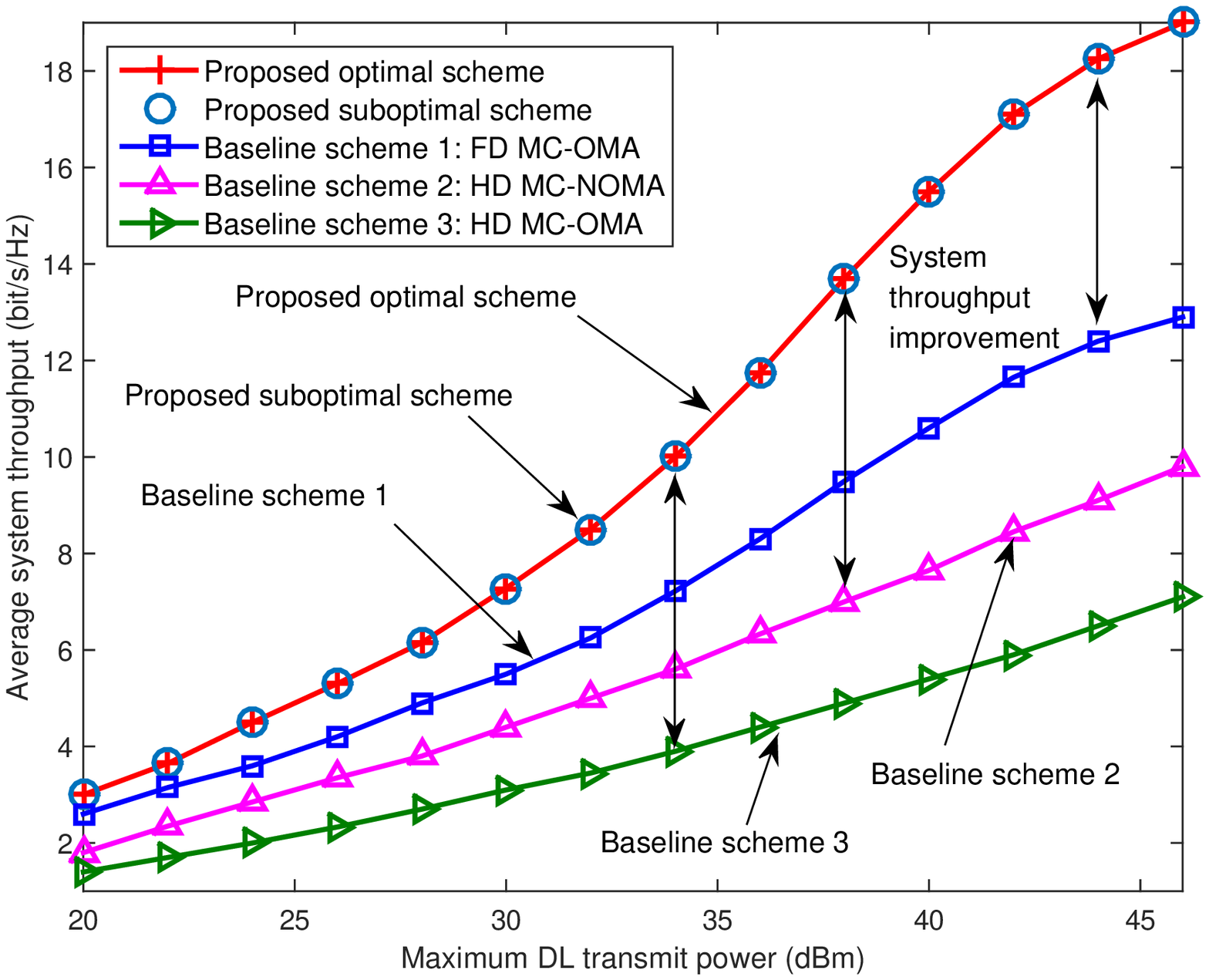}\vspace*{-5mm}
\caption{Average system throughput (bits/s/Hz) versus the maximum DL transmit power at the FD BS, $P_{\mathrm{max}}^{\mathrm{DL}}$ (dBm), for $K=20$ and $J=20$. }
\label{fig:throughput_vs_power}\vspace*{-8mm}
\end{figure}

In Figure \ref{fig:throughput_vs_power}, we investigate the average system throughput versus the maximum DL transmit power at the FD BS, $P_{\mathrm{max}}^{\mathrm{DL}}$, for $K=20$ DL users and $J=20$ UL users.
As can be observed from Figure \ref{fig:throughput_vs_power}, the average system throughput increases monotonically with the maximum DL transmit power $P_{\mathrm{max}}^{\mathrm{DL}}$.
This is because the received  SINR at the DL users can be improved by optimally allocating additional available transmit power  via the solution of the problem in \eqref{pro} which leads to an improvement of the system throughput.
However, there is a diminishing return in the average system throughput when $P_{\mathrm{max}}^{\mathrm{DL}}$ is higher than $40$ dBm. In fact, as the DL transmit power increases, the SI becomes more severe, which degrades the received UL signals.
As a result, the throughput of the UL transmission will decrease and the reduction in  UL throughput  partially neutralizes the improvement in  DL throughput facilitated by the higher DL transmit power. Besides, it can be observed from Figure \ref{fig:throughput_vs_power} that the proposed suboptimal algorithm closely approaches the performance of the proposed optimal power and subcarrier allocation scheme.

Figure \ref{fig:throughput_vs_power} also shows that the average system throughputs of all considered baseline schemes are substantially lower than those of the proposed optimal and suboptimal schemes. In particular, baseline scheme $1$ achieves a lower average system throughput compared to the proposed optimal schemes since  OMA is employed for simultaneous DL and UL transmission and hence, the spectral resource is underutilized. For baseline scheme $2$, DL and UL transmission are separated orthogonally in the time domain which leads to a significant loss in spectral efficiency. Baseline scheme $3$ has the lowest spectral efficiency due to the orthogonal radio resource assignment in both time and frequency.
For the case of $P_{\mathrm{max}}^{\mathrm{DL}}=46$ dBm, the proposed optimal and suboptimal schemes achieve roughly a $49\%$, $188\%$, and $251\%$ higher average system throughput than baseline schemes $1$, $2$, and $3$, respectively.
Besides, the proposed optimal and suboptimal schemes utilize the available transmit power efficiently. In particular, it can be observed from Figure \ref{fig:throughput_vs_power} that for a target system throughput of $6 \text{ bit/s/Hz}$, the proposed schemes enable power reductions of more than $4$ dB, $8$ dB, and $12$ dB compared to baseline schemes $1$, $2$, and $3$, respectively.

\vspace*{-4mm}
\subsection{Average System Throughput versus Total Number of Users}
\begin{figure}
\centering\vspace*{-5mm}
\includegraphics[width=4.3in]{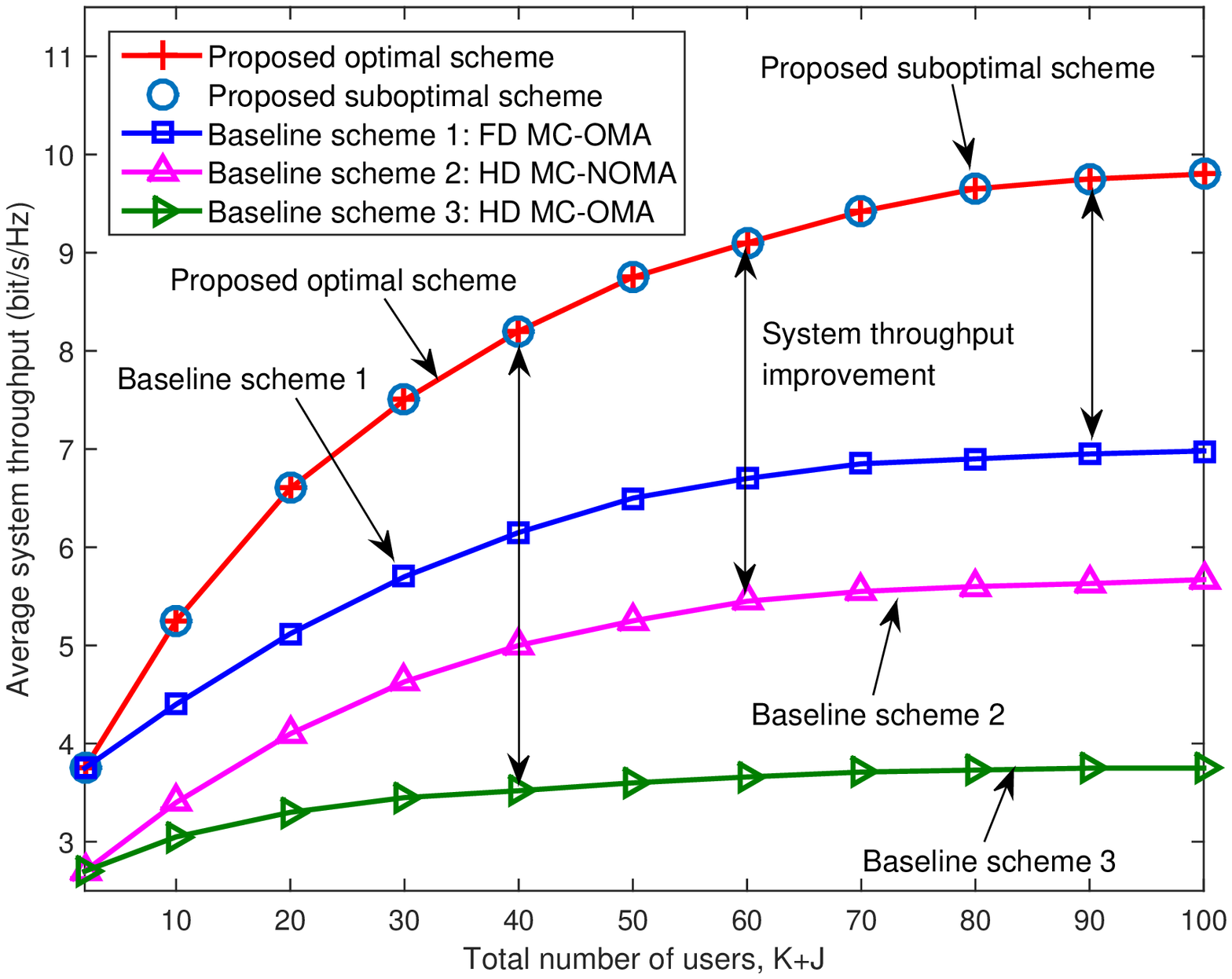}\vspace*{-5mm}
\caption{Average system throughput (bits/s/Hz) versus the total number of users, $K+J$, for $P_{\mathrm{max}}^{\mathrm{DL}}=32 \text{ dBm}$.}
\label{fig:throughput_vs_users}\vspace*{-8mm}
\end{figure}

In Figure \ref{fig:throughput_vs_users}, we investigate the average system throughput versus the total number of users for a maximum transmit power of $P_{\mathrm{max}}^{\mathrm{DL}}=32 \text{ dBm}$. We assume that the numbers of DL and UL users are identical, i.e., $K=J$. As can be observed, the average system throughput for the proposed optimal/suboptimal schemes and the baseline schemes increases with the total number of users since all considered schemes are able to exploit multiuser diversity. However, Figure \ref{fig:throughput_vs_users} also shows that the average system throughput of the proposed optimal and suboptimal schemes grows faster with increasing number of users than that of all the baseline schemes. In fact, the proposed FD MC-NOMA scheme exploits not only the frequency domain but also the power domain for multiple access. Therefore, more degrees of freedom are available in FD MC-NOMA for user selection and power allocation. Thus, the proposed scheme achieves a higher system throughput than the FD MC-OMA system in baseline scheme $1$. Compared to baseline scheme $2$, the proposed optimal scheme always achieves a higher system throughput since it fully utilizes the spectral resource by performing DL and UL communication simultaneously. The traditional HD MC-OMA system in baseline scheme $3$ achieves the lowest average system throughput due its inefficient utilization of the radio resources and diversity. We note that the proposed suboptimal scheme achieves a similar performance as the proposed optimal scheme, even for a relatively large number of users.

\vspace*{-4mm}
\subsection{Average Number of Scheduled Users versus Total Number of Users}
\begin{figure}
\centering\vspace*{-5mm}
\includegraphics[width=4.3in]{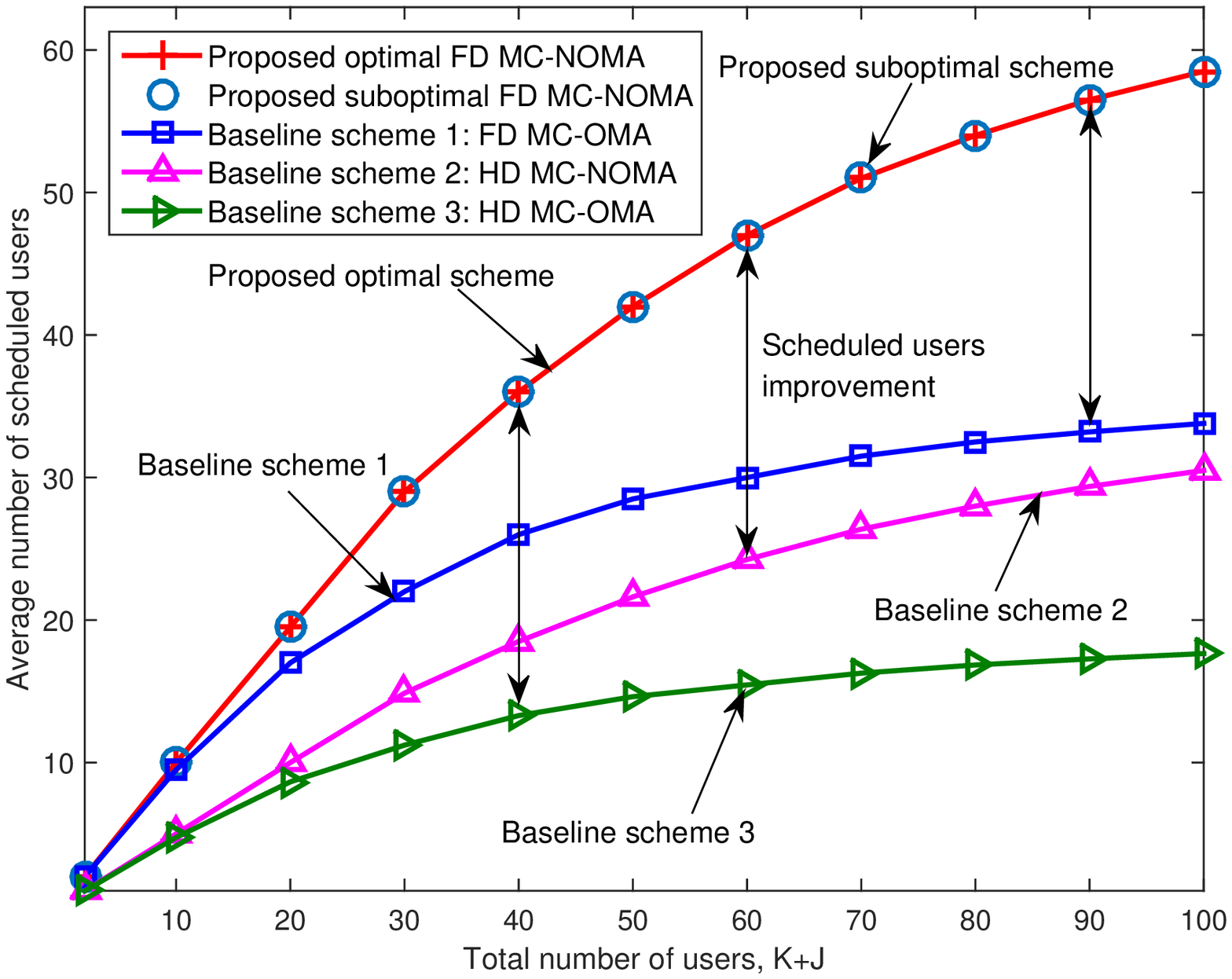}\vspace*{-5mm}
\caption{The average number of scheduled users versus the total number of users, $K+J$, for $P_{\mathrm{max}}^{\mathrm{DL}}=45 \text{ dBm}$. }
\label{fig:scheduled_vs_totalusers}\vspace*{-8mm}
\end{figure}

In Figure \ref{fig:scheduled_vs_totalusers}, we investigate the average number of scheduled users versus the total number of users, $K+J$, for a maximum transmit power of $P_{\mathrm{max}}^{\mathrm{DL}}=45 \text{ dBm}$. Here, the average number of scheduled users is defined as the average number of simultaneously scheduled users in a transmission interval. We further assume that the numbers of DL and UL users are identical.
As can be observed from Figure \ref{fig:scheduled_vs_totalusers}, the number of scheduled users increase monotonically with the total number of users for the proposed optimal/suboptimal schemes and all baseline schemes. However, this increasing trend becomes slower as the total number of users becomes larger since users who suffer from very poor transmission channel conditions may not be allocated any system resources for the maximization of the weighted sum throughput.
Besides, the proposed optimal scheme can accommodate more users compared to baseline schemes $1$ and $3$ since with NOMA more users can be multiplexed on each subcarrier.
In addition, the average number of scheduled users for the proposed schemes is larger than that of baseline scheme $2$ since the proposed schemes can serve DL and UL users simultaneously.
In particular, for the case when $K+J=100$ users are in the system, the proposed schemes provide communication service to $70\%$, $96\%$, and $226\%$ more users than baseline schemes $1$, $2$, and $3$, respectively.
We also note that the proposed suboptimal scheme can accommodate the same average number of scheduled users as the proposed optimal scheme, even when the total number of users is relatively large.

\vspace*{-4mm}
\subsection{Fairness versus Total Number of Users}
\begin{figure}
\centering\vspace*{-5mm}
\includegraphics[width=4.3in]{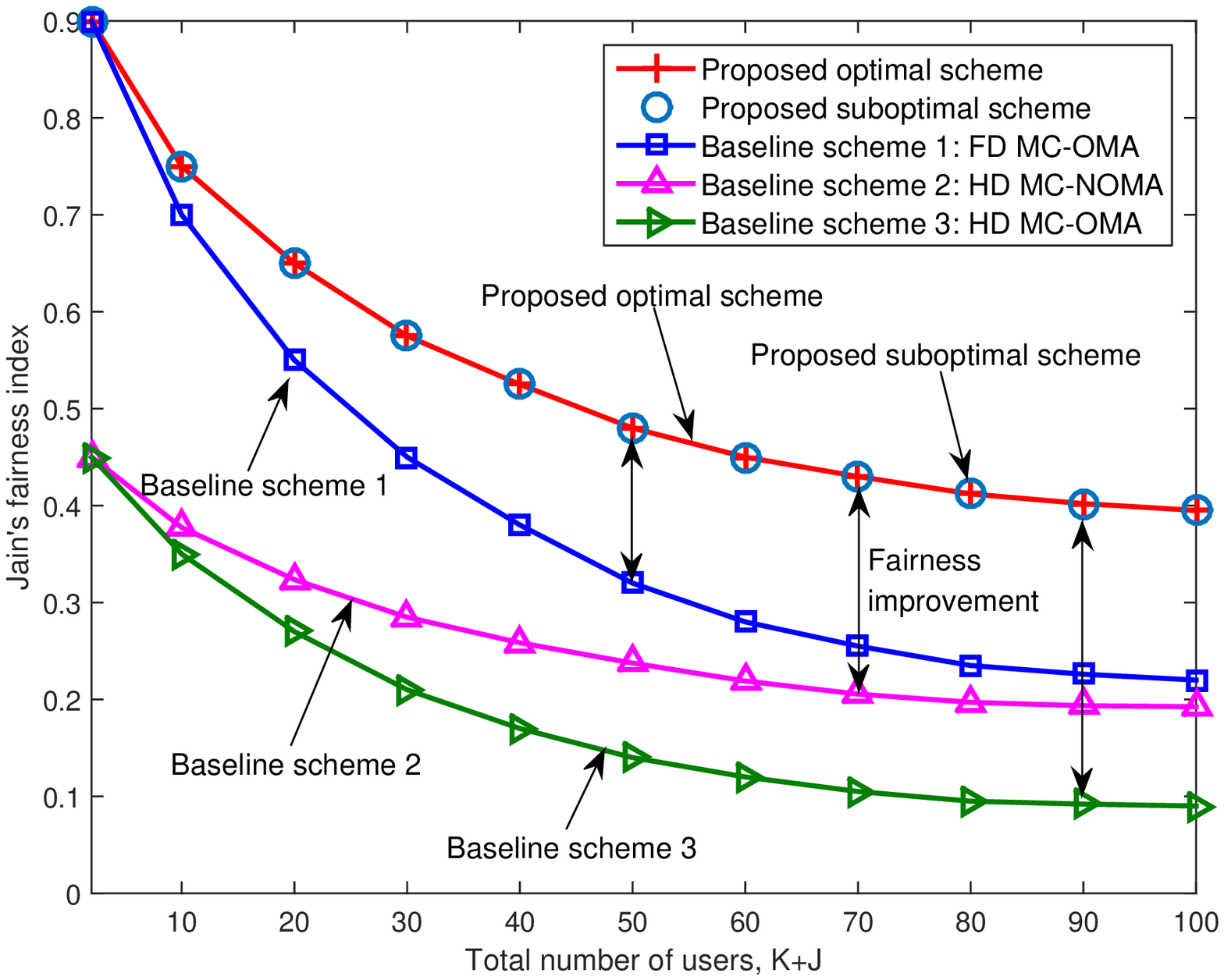}\vspace*{-5mm}
\caption{Jain's fairness index versus the total number of users, $K+J$, for $P_{\mathrm{max}}^{\mathrm{DL}}=45 \text{ dBm}$. }
\label{fig:fairness_vs_user}\vspace*{-8mm}
\end{figure}

In Figure \ref{fig:fairness_vs_user}, we investigate the resource allocation fairness versus the total number of users in the system, for a maximum transmit power of $P_{\mathrm{max}}^{\mathrm{DL}}=45 \text{ dBm}$ at the FD BS. To evaluate the fairness of the proposed schemes and the baseline schemes, we adopt Jain's fairness index \cite{jain1984quantitative} as a performance metric to quantify the notion of fairness. In particular, Jain's fairness index is calculated as $\frac{(\sum_{m=1}^{K}{R_m^{\mathrm{DL}}}+\sum_{r=1}^{J}{R_r^{\mathrm{UL}}})^2}{(K+J)\big(\sum_{m=1}^{K}{R_m^{\mathrm{DL}}}^2 + \sum_{r=1}^{J}{R_r^{\mathrm{UL}}}^2\big)}$, where $R_m^{\mathrm{DL}}$ and $R_r^{\mathrm{UL}}$ are the throughputs of DL user $m$ and UL user $r$, respectively. Jain's fairness index is a real value in the range from $0$ to $1$ and the fairest resource allocation strategy is obtained when Jain's fairness index is equal to $1$ which means every user enjoys the same throughput. In Figure \ref{fig:fairness_vs_user}, it can be observed that the fairness indices achieved by the proposed schemes and the baseline schemes decrease with the total number of users. In fact, the competition among users becomes more fierce when there are more users in the system. In particular, there may be more users with poor channel conditions and lower priorities which may not get access to the communication service. This can also be inferred from Figure \ref{fig:scheduled_vs_totalusers} as the number of scheduled users increases sub-linearly with the number of users for all considered schemes.
Besides, we observe from Figure \ref{fig:fairness_vs_user} that the proposed schemes achieve a higher fairness index compared to baseline schemes $1$ and $3$. The reason behind this is that NOMA, which is adopted in the proposed schemes, is capable of multiplexing more users on each subcarrier compared to baseline schemes $1$ and $3$, respectively, which increases the utilization of multiuser diversity.
We also note that the proposed schemes achieve a higher fairness index compared to baseline scheme $2$. This is because the proposed schemes can provide simultaneous DL and UL communication service which enables a more evenly distributed per user throughput compared to baseline scheme $2$.
In addition, we note that the proposed suboptimal scheme achieves the same fairness index as the proposed optimal scheme, even for  large numbers of users.

\vspace*{-4mm}
\subsection{Average System Throughput versus SI Cancellation Constant}
\begin{figure}
\centering\vspace*{-5mm}
\includegraphics[width=4.3in]{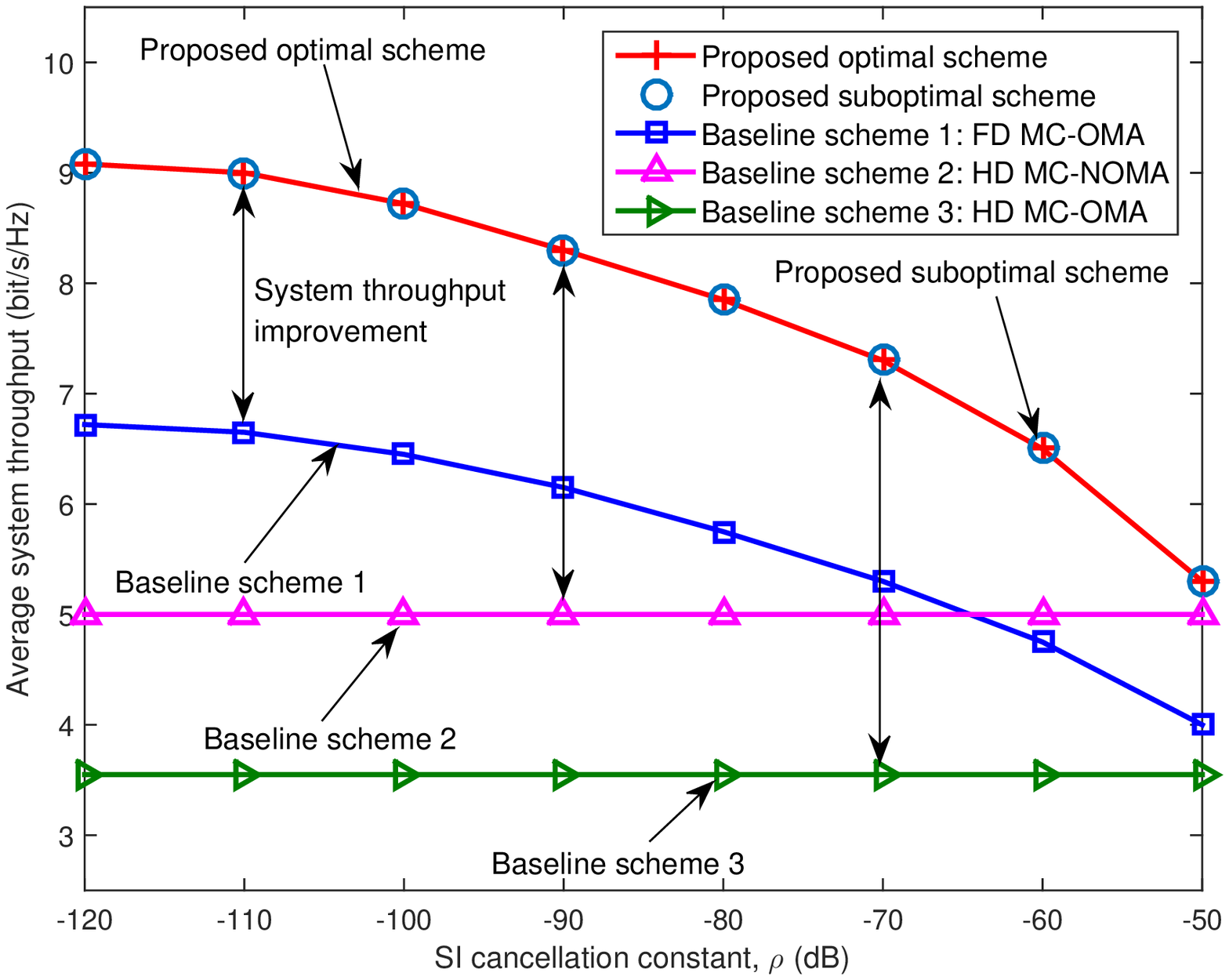}\vspace*{-5mm}
\caption{Average system throughput versus SI cancellation constant, $\rho$ (dB), for $K=20$ DL users, $J=20$ UL users, and $P_{\mathrm{max}}^{\mathrm{DL}}=32 \text{ dBm}$.}
\label{fig:throughput_vs_SI}\vspace*{-8mm}
\end{figure}

In Figure \ref{fig:throughput_vs_SI}, we investigate the average system throughput versus the SI cancellation constant, for $K=20$ DL users, $J=20$ UL users, and a maximum transmit power of $P_{\mathrm{max}}^{\mathrm{DL}}=32 \text{ dBm}$ at the FD BS. As can be observed in Figure \ref{fig:throughput_vs_SI}, the average system throughputs of the proposed schemes and baseline scheme $1$ are monotonically decreasing for increasing SI cancellation constant, $\rho$. This is because a larger SI cancellation constant, $\rho$, causes more residual SI at the FD BS. Thus, the UL reception is more severely impaired by the residual SI at the FD BS which degrades the average throughput for UL transmission. Besides, the average system throughput of the proposed schemes decreases more rapidly than that of baseline scheme $1$. Since, the proposed schemes enable the multiplexing of more UL users on each subcarrier compared to baseline scheme $1$, they are more sensitive to SI. In addition, we notice that the decreasing trend accelerates when $\rho$ exceeds $-110 \text{ dB}$ since then the SI becomes the fundamental system performance bottleneck. We note that the achievability of values of $\rho=-110 \text{ dB}$ has been demonstrated in \cite{FDRad}.
On the other hand, the performances of baseline schemes $2$ and $3$ are independent of the SI cancellation constant due to the adopted HD protocol.

\vspace*{-2mm}
\section{Conclusions}
In this paper, we studied the resource allocation algorithm design for an MC-NOMA system with an FD BS. The algorithm design was formulated as a mixed combinatorial non-convex optimization problem for the maximization of the weighted sum throughput of the system. Monotonic optimization was applied for solving the proposed optimization problem optimally. The resulting optimal power and subcarrier allocation policy served as a performance benchmark due to its high computational complexity.
Therefore, a suboptimal algorithm based on successive convex approximation was also proposed to strike a balance between computational complexity and optimality.
Simulation results showed that the proposed suboptimal algorithm obtained a close-to-optimal performance in a small number of iterations.
Besides, our results revealed that a substantial improvement of system throughput can be achieved by employing the proposed FD MC-NOMA scheme compared to baseline schemes employing FD MC-OMA, HD MC-NOMA, and HD MC-OMA. Furthermore, the proposed FD MC-NOMA scheme was shown to provide a good balance between improving the system throughput and maintaining fairness among users.

\vspace*{-1mm}
\section*{Appendix-Proof of Theorem 1}
We start the proof by utilizing the \emph{abstract Lagrangian duality} \cite{che2014Joint,goh2002duality,rockafellar1974conjugate}.
We first define
\begin{eqnarray}
&&\hspace*{-3mm}{\cal{L}}(\tilde{\mathbf{p}},\tilde{\mathbf{q}},\mathbf{s},\eta) \\
=&&\hspace*{-3mm}\sum_{i=1}^{{N_{\mathrm{F}}}}\sum_{m=1}^{K} \sum_{n=1}^{K} \overset{J}{\underset{r=1}{\sum}} \overset{J}{\underset{t=1}{\sum}} -U_{m,n,r,t}^i(\tilde{\mathbf{p}},\tilde{\mathbf{q}}) + \eta\Big(\overset{{N_{\mathrm{F}}}}{\underset{i=1}{\sum}} \overset{K}{\underset{m=1}{\sum}} \overset{K}{\underset{n=1}{\sum}} \overset{J}{\underset{r=1}{\sum}} \overset{J}{\underset{t=1}{\sum}} s_{m,n,r,t}^i - (s_{m,n,r,t}^i)^2\Big). \notag
\end{eqnarray}
Then, we note that \vspace*{-2mm}
\begin{eqnarray}
\hspace*{-13mm}&&\hspace*{-3mm} \underset{\eta \ge 0}{\maxo} \quad {\cal {L}}(\tilde{\mathbf{p}},\tilde{\mathbf{q}},\mathbf{s},\eta)  \notag \\
\hspace*{-13mm}&=&\hspace*{-3mm} \begin{cases}
\hspace*{-0mm} \overset{N_{\mathrm{F}}}{\underset{i=1}{\sum}} \overset{K}{\underset{m=1}{\sum}} \overset{K}{\underset{n=1}{\sum}} \overset{J}{\underset{r=1}{\sum}} \overset{J}{\underset{t=1}{\sum}} -U_{m,n,r,t}^i(\tilde{\mathbf{p}},\tilde{\mathbf{q}}),  &   \overset{{N_{\mathrm{F}}}}{\underset{i=1}{\sum}} \overset{K}{\underset{m=1}{\sum}} \overset{K}{\underset{n=1}{\sum}} \overset{J}{\underset{r=1}{\sum}} \overset{J}{\underset{t=1}{\sum}} s_{m,n,r,t}^i - (s_{m,n,r,t}^i)^2 \le 0, \\
\hspace*{-0mm} \infty, & \text{otherwise}.
\end{cases}
\end{eqnarray}
Therefore, the optimization problem in \eqref{subopt-pro} can be equivalently written as
\begin{eqnarray} \label{eqv-subopt-pro}
d^* = \underset{\tilde{\mathbf{p}},\tilde{\mathbf{q}},\mathbf{p},\mathbf{q},\mathbf{s} \in \mathbf{\Omega}}{\mino} \quad \underset{\eta \ge 0}{\maxo} \quad {\cal {L}}(\tilde{\mathbf{p}},\tilde{\mathbf{q}},\mathbf{s},\eta),
\end{eqnarray}
where $\mathbf{\Omega}$ is the feasible set spanned by constraints $\mbox{C1--C3}, \text{C4}\mbox{b},$ and $\mbox{C5--C15}$, and $d^*$ is the optimal value of  optimization problem \eqref{subopt-pro}. On the other hand, the dual problem of \eqref{subopt-pro} is
\begin{eqnarray} \label{dual-subopt-pro}
\underset{\eta \ge 0}{\maxo} \,\, \underset{\tilde{\mathbf{p}},\tilde{\mathbf{q}},\mathbf{p},\mathbf{q},\mathbf{s} \in \mathbf{\Omega}}{\mino} \,\,   {\cal {L}}(\tilde{\mathbf{p}},\tilde{\mathbf{q}},\mathbf{s},\eta)
\,\,\, = \,\,\, \underset{\eta \ge 0}{\maxo} \,\, \Theta (\eta),
\end{eqnarray}
where $\Theta (\eta)$ is defined as $\Theta (\eta) \triangleq \underset{\tilde{\mathbf{p}},\tilde{\mathbf{q}},\mathbf{p},\mathbf{q},\mathbf{s} \in \mathbf{\Omega}}{\mino} \quad   {\cal {L}}(\tilde{\mathbf{p}},\tilde{\mathbf{q}},\mathbf{s},\eta)$ for notational simplicity.
Then, the equivalent primal problem \eqref{eqv-subopt-pro} and dual problem \eqref{dual-subopt-pro} meet the following inequalities:
\begin{eqnarray} \label{eta-uper-bound}
\underset{\eta \ge 0}{\maxo} \quad \Theta (\eta) &=& \underset{\eta \ge 0}{\maxo} \quad \underset{\tilde{\mathbf{p}},\tilde{\mathbf{q}},\mathbf{p},\mathbf{q},\mathbf{s} \in \mathbf{\Omega}}{\mino} \quad   {\cal {L}}(\tilde{\mathbf{p}},\tilde{\mathbf{q}},\mathbf{s},\eta) \notag \\
&\overset{(a)}{\le}& \underset{\tilde{\mathbf{p}},\tilde{\mathbf{q}},\mathbf{p},\mathbf{q},\mathbf{s} \in \mathbf{\Omega}}{\mino} \quad \underset{\eta \ge 0}{\maxo} \quad {\cal {L}}(\tilde{\mathbf{p}},\tilde{\mathbf{q}},\mathbf{s},\eta) = d^*,
\end{eqnarray}
where $(a)$ is due to the weak duality \cite{book:convex}. We note that ${\cal{L}}(\tilde{\mathbf{p}},\tilde{\mathbf{q}},\mathbf{s},\eta)$ is a monotonically increasing function in variable $\eta$ since $\overset{{N_{\mathrm{F}}}}{\underset{i=1}{\sum}} \overset{K}{\underset{m=1}{\sum}} \overset{K}{\underset{n=1}{\sum}} \overset{J}{\underset{r=1}{\sum}} \overset{J}{\underset{t=1}{\sum}} s_{m,n,r,t}^i - (s_{m,n,r,t}^i)^2 \ge 0$ for $\mathbf{s} \in \mathbf{\Omega}$. As a result, $\Theta (\eta)$ is an increasing function with $\eta$. Besides, \eqref{eta-uper-bound} implies that $\Theta (\eta)$ is bounded from above by the optimal value of \eqref{subopt-pro}, i.e., $d^*$.
We suppose that the optimal solution of the dual problem in \eqref{dual-subopt-pro} is denoted as $\widehat{\eta}^*$ and $\widehat{\mathbf{\Xi}}^* \triangleq \big \{\widehat{\tilde{\mathbf{p}}}^*,\widehat{\tilde{\mathbf{q}}}^*,\widehat{\mathbf{p}}^*,\widehat{\mathbf{q}}^*,\widehat{\mathbf{s}}^* \big \}$. Then, we study the solution structure of the dual problem \eqref{dual-subopt-pro} by considering the following two cases.
For the first case, we assume that $\overset{{N_{\mathrm{F}}}}{\underset{i=1}{\sum}} \overset{K}{\underset{m=1}{\sum}} \overset{K}{\underset{n=1}{\sum}} \overset{J}{\underset{r=1}{\sum}} \overset{J}{\underset{t=1}{\sum}} \widehat{s}_{m,n,r,t}^{i*} - (\widehat{s}_{m,n,r,t}^{i*})^2 = 0$ for the dual problem in \eqref{dual-subopt-pro}, where $\widehat{s}_{m,n,r,t}^{i*}$ are the elements of $\widehat{\mathbf{s}}^*$. As a result, $\widehat{\mathbf{\Xi}}^*$ is also a feasible solution to primal problem in \eqref{subopt-pro}. Consequently, by substituting $\widehat{\mathbf{\Xi}}^*$ into the optimization problem in \eqref{subopt-pro}, we have
\begin{eqnarray} \label{eta-lower-bound}
d^* \le \sum_{i=1}^{{N_{\mathrm{F}}}}\sum_{m=1}^{K} \sum_{n=1}^{K} \overset{J}{\underset{r=1}{\sum}} \overset{J}{\underset{t=1}{\sum}} -U_{m,n,r,t}^i(\widehat{\tilde{\mathbf{p}}}^*,\widehat{\tilde{\mathbf{q}}}^*) \overset{(b)}{=}{\cal{L}}(\widehat{\tilde{\mathbf{p}}}^*,\widehat{\tilde{\mathbf{q}}}^*,\widehat{\mathbf{s}}^*,\widehat{\eta}^*)=\Theta (\widehat{\eta}^*),
\end{eqnarray}
where $(b)$ is due to the assumption of $\overset{{N_{\mathrm{F}}}}{\underset{i=1}{\sum}} \overset{K}{\underset{m=1}{\sum}} \overset{K}{\underset{n=1}{\sum}} \overset{J}{\underset{r=1}{\sum}} \overset{J}{\underset{t=1}{\sum}} \widehat{s}_{m,n,r,t}^{i*} - (\widehat{s}_{m,n,r,t}^{i*})^2 = 0$. By combining \eqref{eta-uper-bound} and \eqref{eta-lower-bound}, we can conclude that the gap  between the equivalent primal problem \eqref{eqv-subopt-pro} and the dual problem \eqref{dual-subopt-pro} is zero, i.e.,
\begin{eqnarray}
\underset{\eta \ge 0}{\maxo} \quad \underset{\tilde{\mathbf{p}},\tilde{\mathbf{q}},\mathbf{p},\mathbf{q},\mathbf{s} \in \mathbf{\Omega}}{\mino} \quad   {\cal {L}}(\tilde{\mathbf{p}},\tilde{\mathbf{q}},\mathbf{s},\eta)
= \underset{\tilde{\mathbf{p}},\tilde{\mathbf{q}},\mathbf{p},\mathbf{q},\mathbf{s} \in \mathbf{\Omega}}{\mino} \quad \underset{\eta \ge 0}{\maxo} \quad {\cal {L}}(\tilde{\mathbf{p}},\tilde{\mathbf{q}},\mathbf{s},\eta)
\end{eqnarray}
must hold for $\overset{{N_{\mathrm{F}}}}{\underset{i=1}{\sum}} \overset{K}{\underset{m=1}{\sum}} \overset{K}{\underset{n=1}{\sum}} \overset{J}{\underset{r=1}{\sum}} \overset{J}{\underset{t=1}{\sum}} s_{m,n,r,t}^i - (s_{m,n,r,t}^i)^2 = 0$. Furthermore, the monotonicity of $\Theta (\eta)$ with respect to $\eta$ implies that
\begin{eqnarray}
\Theta (\eta) =d^*, \,\, \forall \eta \ge \widehat{\eta}^*,
\end{eqnarray}
which confirms the result of Theorem \ref{thm-penalty}.

Then, we study the case of $\overset{{N_{\mathrm{F}}}}{\underset{i=1}{\sum}} \overset{K}{\underset{m=1}{\sum}} \overset{K}{\underset{n=1}{\sum}} \overset{J}{\underset{r=1}{\sum}} \overset{J}{\underset{t=1}{\sum}} \widehat{s}_{m,n,r,t}^{i*} - (\widehat{s}_{m,n,r,t}^{i*})^2 > 0$ for the dual problem in \eqref{dual-subopt-pro}. In this case, $\Theta (\widehat{\eta}^*)=\underset{\eta \ge 0}{\maxo} \,\, \Theta (\eta) \rightarrow \infty$ is unbounded from above since the function $\Theta (\eta)$ is monotonically increasing in $\eta$. This contradicts the inequality in \eqref{eta-uper-bound} as the primal problem in \eqref{subopt-pro} has a finite objective value. Therefore,  $\overset{{N_{\mathrm{F}}}}{\underset{i=1}{\sum}} \overset{K}{\underset{m=1}{\sum}} \overset{K}{\underset{n=1}{\sum}} \overset{J}{\underset{r=1}{\sum}} \overset{J}{\underset{t=1}{\sum}} \widehat{s}_{m,n,r,t}^{i*} - (\widehat{s}_{m,n,r,t}^{i*})^2 = 0$ holds for the optimal solution and the proof of Theorem \ref{thm-penalty} is completed.

\vspace*{-0mm}


\end{document}